# Vector Properties of Entanglement in a Three-Qubit System

Dmitry B. Uskov [1,2*], Paul M. Alsing [3]

[1]*Department of Mathematics and Natural Sciences, Brescia University, Owensboro, Kentucky 42301, USA*
[2]*Department of Physics and Engineering Physics, Tulane University, New Orleans, Louisiana 70118, USA*
[3]*Air Force Research Laboratory, Information Directorate, Rome, New York 13441, USA*

**Abstract**: We suggest a dynamical vector model of entanglement in a three qubit system based on isomorphism between $su(4)$ and $so(6)$ Lie algebras. This model allows one to write an evolution equation for 3-qubit entanglement parameters under an arbitrary pairwise qubit coupling. Generalizing Plücker-type description of three-qubit local invariants we introduce three pairs of real-valued $3D$ vectors (denoted here as $\mathbf{A}_{R,I}$, $\mathbf{B}_{R,I}$ and $\mathbf{C}_{R,I}$). Magnitudes of these vectors determine two- and three-qubit entanglement parameters of the system. We show that evolution of vectors $\mathbf{A}$, $\mathbf{B}$ and $\mathbf{C}$ under local $SU(2)$ operations is identical to $SO(3)$ evolution of single-qubit Bloch vectors of qubits a, b and c correspondingly. At the same time, general two-qubit $su(4)$ Hamiltonians incorporating a-b, a-c and b-c two-qubit coupling terms generate $SO(6)$ coupling between vectors $\mathbf{A}$ and $\mathbf{B}$, $\mathbf{A}$ and $\mathbf{C}$, and $\mathbf{B}$ and $\mathbf{C}$ correspondingly. It turns out that dynamics of entanglement induced by different two-qubit coupling terms is entirely determined by mutual orientation of vectors $\mathbf{A}$, $\mathbf{B}$, $\mathbf{C}$. We illustrate the power of this vector description of entanglement by solving quantum control problems involving transformations between $W$, Greenberg-Horne-Zeilinger ($GHZ$) and biseparable states.

\* Corresponding author: dmitry.uskov@brescia.edu

# I. Introduction

Generation, control and characterization of entanglement of multipartite quantum states is one of the foci of ongoing theoretical and experimental research in quantum information processing [1-3]. Three-qubit entangled state, such as $GHZ$, $W$ and various three-qubit cluster states has been generated with trapped ions [4] and photons [5, 6]. However theoretical description of dynamics and control of multi-partite entanglement is practically absent in the literature.

Previous work dedicated to entanglement in quantum information focused on static characteristics of entanglement (e.g. local invariants [1-3]). In contrast, the model presented here provides insights into the dynamics of entanglement induced by qubit-qubit coupling.

Quantitative description of multipartite entanglement of pure three-qubit systems is based on the idea of a $\mathbb{CP}^7 / SU(2)^{\otimes 3}$ quotient space: two three-qubit quantum states belong to the same equivalence class if they are related by a set of local single-qubit transformations ($\mathbb{CP}^7$ space here is simply a space of normalized three-qubit states equivalent up to a global $U(1)$ phase multiplication). It has been identified that there are, in general, five independent invariants which can be used as coordinates on such a space [7-14]. Physically important insight into the properties of three-qubit entanglement has been established in [15, 16] where three-tangle was identified as an important physical parameter characterizing genuine three-body entanglement. This parameter is related to two-body entanglement via so-called Coffman-Kundu-Wootters (CKW) inequality [15].

Further progress in clarifying physical meaning of three-body entanglement was achieved by Peter Lévay in a series of papers where he used mathematical ideas of fiber bundle theory [17-20]. This approach revealed an alternative geometric way of quantifying entanglement whereby three-qubit states are projected onto Klein quadratic embedded in $CP^5$ space [18]. In the context of this approach a set of six Plücker coordinates on Grassmannian $G(4,2,\mathbb{C})$ manifold [21] appears as a natural algebraic tool allowing to factor out action of single-qubit local operations in a consistent geometric form (see also recent paper [22]). In the paper on three-qubit entanglement [18] Peter Lévay used isomorphic mapping of $SL(2) \otimes SL(2)$ group onto $SO(4,\mathbb{C})$ group (this technique first appeared in the context of classification of two-qubit entangling operations in the form of $SU(2) \otimes SU(2) = SO(4)$ isomorphism [23, 24]). However, this approach is limited to only local operations providing little insight into the dynamics of entanglement under qubit-qubit coupling. $SU(2) \otimes SU(2) = SO(4)$ isomorphism involves only single qubit transformations. It is embedded in a larger $SU(6)/\mathbb{Z}_2 = SO(6)$ isomorphism (which is at the foundation of our work). The crucial difference is that the latter includes qibit-qubit coupling operators while the former contains only local qubit transformations. In the present paper we address the following problem: how does entanglement change under the action of non-local set of transformations including qubit-qubit coupling terms?

Since a general two-qubit group is the set of $SU(4)$ operations, action of this group on six Plücker coordinates generates a six-dimensional representation of $SU(4)$ isomorphic to $SO(6)$ group of real orthogonal rotations. However, the situation is little bit more complicated than standard abstract description of $su(4) = so(6)$ Lie algebra isomorphism [25]: i) $su(4)$ action on Plücker coordinates does not manifest itself in a canonical form of $so(6)$ real-valued $6\times 6$ antisymmetric generators, ii) six-dimensional space of Plücker coordinates is a complex six-dimensional space $\mathsf{C}^6 \equiv \mathsf{R}^6 \oplus \mathsf{R}^6$ i.e. a direct sum of two real six-dimensional vectors spaces (denoted as $\mathsf{R}^6$).

As mentioned above, original Plücker variables need to be modified by a $U(6)$ transformation (see details in Section IV) in order to obtain a canonical representation of $SO(6)$ in the form of real-valued $6\times 6$ antisymmetric generators. This operation results in a modification of original Plücker variables; here we refer to a new set of variables as quantum Plücker variables, or simply $\mathbf{q}$-variables (or $\mathbf{q}$-vectors). These new coordinates have a set of interesting useful properties. By construction original Plücker variables are explicitly partition-dependent. However, there exist trivial relations between $\mathbf{q}$-vectors characterizing three different partitions (a(bc), c(ab) and b(ca), which we will also label as partitions (1), (2) and (3), respectively). These relations immediately allows us to reduce redundant $18 = 6\times 3$ Plücker complex parameters describing all three partitions to only three complex three-dimensional vectors (we call them $\mathbf{A}$, $\mathbf{B}$ and $\mathbf{C}$ in order to reflect their relation to qubits a, b and c correspondingly). Partition-independent description of three-qubit entanglement requires only three pairs of real three-dimensional vectors $\mathbf{A}_R = \mathrm{Re}(\mathbf{A})$, $\mathbf{A}_I = \mathrm{Im}(\mathbf{A})$, $\mathbf{B}_R = \mathrm{Re}(\mathbf{B})$, $\mathbf{B}_I = \mathrm{Im}(\mathbf{B})$, $\mathbf{C}_R = \mathrm{Re}(\mathbf{C})$, $\mathbf{C}_I = \mathrm{Im}(\mathbf{C})$. Interestingly, these vectors obey the same set of dynamic equations as three single-qubit Bloch vectors (details are presented in section IV below).

Since two- and three-tangles are related to magnitudes of vectors $\mathbf{A}$, $\mathbf{B}$, $\mathbf{C}$ in a very trivial algebraic fashion these vectors provide transparent geometric description of entanglement dynamics under two-qubit coupling making quantum Plücker description of three-qubit system a very useful tool for solving quantum control problems. Geometric operations (rotations) of vectors $\mathbf{A}$, $\mathbf{B}$, $\mathbf{C}$ can be easily tailored to achieve a desired goal of transforming one state to another state. Due to existence of explicit algebraic relation between $SO(6)$ and $SU(4)$ rotations [26, 27] we can establish a protocol when desired quantum state transformation is visualized and constructed and as two-stage set of rotations: 3D Bloch-type rotations of vectors $\mathbf{A}$, $\mathbf{B}$, $\mathbf{C}$ followed by a set of qubit-qubit couplings which take the form of couplings between these three vectors. As soon as appropriate rotations and couplings of vectors $\mathbf{A}$, $\mathbf{B}$, $\mathbf{C}$ are established using geometric considerations, one can immediately derive a set of corresponding $SU(4)$ quantum operations which physically implement desired transformation of a quantum state.

The outline of this paper is as follows. In Section II we define the Plücker variables for pure tripartite quantum states and list some of their properties. In Section III we discuss the important accidental Lie group isomorphism between local two-qubit operations and orthogonal transformations of complex four-vectors. We review Lévay's [18] derivation of the three- and two-tangles as invariants of a six-component Plücker vector. In Section IV we introduce a modification of the Plücker vector that will allow us to efficiently investigate two-qubit entangling operations. We derive the evolution of this new Plücker vector (which we call $\mathbf{q}$-vector) under a general two-qubit coupling Hamiltonian. In section V we show how the $\mathbf{q}$-vectors for each partitioning of the three qubits behave like Bloch vectors. In Section VI, we show the relationship of the $\mathbf{q}$-vectors in each qubit partitioning to each other, and show that systems reduce down to three complex Bloch-like three-vectors $\mathbf{A}$, $\mathbf{B}$, $\mathbf{C}$. In Section VII, we show how the three- and two-tangles are related to invariants of $\mathbf{A}$, $\mathbf{B}$, $\mathbf{C}$. After the previous mathematical reduction of three-qubit operations as transformations of the relevant Bloch-like vectors $\mathbf{A}$, $\mathbf{B}$, $\mathbf{C}$, in Section VIII we show how non-intuitive entanglement control and manipulation in quantum state space is greatly facilitated by intuitive operations in $\mathbf{q}$-space. As nontrivial examples we show how one can transform a $W$ state into a $GHZ$ state, a biseparable state into a $GHZ$ state and $W$ state into a biseparable state. In Section IX we present a discussion of our work and indicate avenues for future research.

## II. Plücker Variables: Definition

Consider a pure three-qubit state

$$|\psi\rangle = \sum_{i,j,k=0,1} c_{ijk} |i,j,k\rangle. \tag{1}$$

Ignoring normalization, eight coefficients $c_{ijk}$ belong to a linear complex eight-dimensional space $\mathbb{C}^8$. Single-qubit operators $\mathbf{V}^{(a,b,c)} \subset SL(2,\mathbb{C})$, acting on qubits $a$, $b$ and $c$ correspondingly, induce transformations of coefficients $c_{ijk} \to c'_{ijk}$ given by

$$c'_{ijk} = \sum_{n=0,1} V^{(a)}_{in} V^{(b)}_{jm} V^{(c)}_{kp} c_{nmp}. \tag{2}$$

In matrix notations

$$\mathbf{C}^{(1,2,3)} \to \mathbf{C}^{(1,2,3)} \left[ \mathbf{V}^{(a,b,c)} \right]^T. \tag{3}$$

Three $4 \times 2$ matrices $\mathbf{C}^{(1,2,3)}$ emerge due to three bipartite arrangements of qubits: $a(bc)$, $b(ca)$ and $c(ab)$. We label these partitions in equation (3) as $1$, $2$ and $3$ (note: for ease of notation, $\mathbf{C}^{(1,2,3)}$ refers collectively to any of matrices $\mathbf{C}^{(1)}$, $\mathbf{C}^{(2)}$ or $\mathbf{C}^{(3)}$ constructed for all three partitions). We will consistently use the following notation rule concerning labeling partitions and qubits: operators and variables referring to a specific partition have partition number as a superscripts; at the same time letter superscripts $(a),(b),(c),(bc),(ca)$ and $(ab)$ denote operators acting on specific qubits or pairs of qubits. For example, operator $\mathbf{V}^{(a)}$ in equation (3) is a $2 \times 2$ matrix of the local operator acting on qubit $(a)$, while $\mathbf{C}^{(1)}$ refers to a $4 \times 2$ matrix arrangement of state coefficients $c_{ijk}$ specific for partition $a(bc)$. This matrix is given by equation

$$\mathbf{C}^{(1)} = \left( \mathbf{c}^{(1)}_0, \mathbf{c}^{(1)}_1 \right), \; \mathbf{c}^{(1)}_0 = (c_{000}, c_{001}, c_{010}, c_{011})^T, \; \mathbf{c}^{(1)}_1 = (c_{100}, c_{101}, c_{110}, c_{111})^T. \tag{4}$$

Matrices $\mathbf{C}^{(2)}$ and $\mathbf{C}^{(3)}$ are generated by cyclic permutations of subscript indexes $ijk \to jki \to kij$ and simultaneous cyclic change of partition number $1 \to 2 \to 3$ in equation (4).

Since $Det(\mathbf{V}^{(a,b,c)})=1$, subdeterminants of matrices $\mathbf{C}^{(1)}$, $\mathbf{C}^{(2)}$ and $\mathbf{C}^{(3)}$, do not change under transformations given by equation (3). These subdeterminants are quadratic polynomial invariants of local transformations acting on qubits a, b and c, correspondingly. However, subdeterminants of $\mathbf{C}^{(1)}$, $\mathbf{C}^{(2)}$ and $\mathbf{C}^{(3)}$ will change under local transformations acting on qubits $b$ or $c$, qubits $c$ or $a$ and qubits $a$ or $b$, correspondingly. In other words, subdeterminants of $\mathbf{C}^{(1)}$, $\mathbf{C}^{(2)}$ and $\mathbf{C}^{(3)}$ generate three sets of six quadratic polynomials; each of these sets is invariant only under corresponding single-qubit group of local operations. All entanglement parameters of three-qubit states, except for Kempe invariant [10], can be expressed in terms of Plücker variables. We will show a natural method of constructing symmetric partition-independent description of three-qubit measures of mutual entanglement.

From a geometric point of view one can associate a set of subdeterminants with Grassmannian manifolds [18]. Subdeterminants of $\mathbf{C}^{(1,2,3)} = \left(\mathbf{c}_0^{(1,2,3)}, \mathbf{c}_1^{(1,2,3)}\right)$ provide homogeneous (Plücker) coordinates on $G(2,4,\mathbb{C})$ manifold. Explicitly, we have

$$\begin{aligned}
\mathcal{P}^{(1)}_{(n,m),(k,l)} &= c_{0,n,m} c_{1,k,l} - c_{1,n,m} c_{0,k,l}, \\
\mathcal{P}^{(2)}_{(n,m),(k,l)} &= c_{m,0,n} c_{l,1,k} - c_{m,1,n} c_{l,0,k}, \\
\mathcal{P}^{(3)}_{(n,m),(k,l)} &= c_{n,m,0} c_{k,l,1} - c_{n,m,1} c_{k,l,0}.
\end{aligned} \qquad (5)$$

To simplify notations we use decimal-binary conversion by introducing indexes $r = 2n+m+1$ and $r' = 2k+l+1$, $r, r' = 1,...,4$, such that

$$\mathrm{P}^{(1,2,3)}_{r,r'} = \mathcal{P}^{(1,2,3)}_{(n,m),(k,l)}. \qquad (6)$$

Since $\mathrm{P}^{(1,2,3)}_{n,m} = -\mathrm{P}^{(1,2,3)}_{m,n}$ each matrix $\mathbf{P}^{(1,2,3)}$ has only six independent parameters which can be rearranged in the form of vectors $\mathbf{p}^{(1,2,3)} \in \mathbb{R}^6$.

$$\mathbf{p}^{(s)} = \left(\mathrm{P}^{(s)}_{1,2}, \mathrm{P}^{(s)}_{1,3}, \mathrm{P}^{(s)}_{1,4}, \mathrm{P}^{(s)}_{2,3}, \mathrm{P}^{(s)}_{2,4}, \mathrm{P}^{(s)}_{3,4}\right)^T, \quad s=1,2,3. \qquad (7)$$

As we know [28] there exists an additional constraint on these variables called Plücker relation. In terms of vectors $\mathbf{p}^{(1,2,3)}$, given by equation (7), this relation can be written as a bilinear quadratic form

$$\mathbf{p}^T \cdot \mathbf{\Omega} \cdot \mathbf{p} = 0. \qquad (8)$$

Here symmetric $6\times 6$ matrix $\mathbf{\Omega}$ has six non-zero anti-diagonal entries

$$\Omega_{1,6} = \Omega_{3,4} = \Omega_{4,3} = \Omega_{6,1} = 1, \quad \Omega_{2,4} = \Omega_{4,2} = -1. \qquad (9)$$

Invariance of Plücker variables under local transformations, including non-unitary $SL(2,\mathbb{C})$ operators, make them instrumental in geometric description of entanglement properties of multi-qubit systems [17-20, 22]. For example, Plücker coordinates provide a link between twistor theory and geometric description of different classes of entanglement [18].

### III. Plücker invariant forms for two-qubit local transformations and $SL(2,\mathbb{C}) \otimes SL(2,\mathbb{C}) \equiv SO(4,\mathbb{C})$ accidental Lie group isomorphism.

Special linear $SL(2)$ groups emerge naturally as an extension of $SU(2)$ group in the context of single qubit operations involving measurements (see, for example, review [2], third paragraph after equation (101)).

To simplify equation in this section we chose a specific partition (1) - $a(bc)$. Equations for partitions (2) - $b(ca)$ and (3) - $c(ab)$ are obtained by trivial cyclic relabeling of qubits.

A key mathematical relation in this section is the isomorphism between Lie Group of local operators $SL(2,\mathbb{C})^{(b)} \otimes SL(2,\mathbb{C})^{(c)}$ acting on qubits $b$ and $c$, and the group of complex orthogonal rotations $SO(4,\mathbb{C})$, acting on vectors $\mathbf{c}_0^{(1)}, \mathbf{c}_1^{(1)} \subset \mathbb{C}^4$ defined by equation (4). To fix notations, Lie algebra $sl(2,\mathbb{C})$ is spanned, as a linear space over real numbers, by a set of six traceless complex matrices, which may be chosen to be $i\sigma_{x,y,z}$ and $\sigma_{x,y,z}$, for example. Skew Hermitian matrices $i\sigma_{x,y,z}$ span compact $su(2)$ subalgebra (also known as Lorentz rotations) of $sl(2,\mathbb{C})$, and the set $\sigma_{x,y,z}$ (boosts) generates Cartan complement $\wp$. Lie algebra $sl(2,\mathbb{C}) = su(2) \oplus \wp$. Commutators $[su, su]$ close in $su$, $[su, \wp]$ close in $\wp$ and $[\wp, \wp]$ close in $su$.

Importance of the isomorphism between two-qubit local unitary transformation and $SO(4)$ Lie Group for classification of two-qubit unitary entangling transformations was established in series of papers [23,24]. At the level of Lie algebras we have $su(2)^{(b)} \oplus su(2)^{(c)} \equiv so(4,\mathbb{R})$. Since complexification of $su(2)$ (sometimes denoted as $su(2)_{\mathbb{C}}$) is apparently isomorphic to $sl(2,\mathbb{C})$ the complexification of $su(2)^{(b)} \oplus su(2)^{(c)}$ is isomorphic to $so(4,\mathbb{C})$, i.e. $sl(2,\mathbb{C})^{(b)} \oplus sl(2,\mathbb{C})^{(c)} \equiv su(2)_{\mathbb{C}}^{(b)} \oplus su(2)_{\mathbb{C}}^{(b)} \equiv so(4,\mathbb{C})$.

The $so(4,\mathbb{C})$ Lie algebra is a 12-dimensional linear space of antisymmetric complex matrices. The isomorphism between $sl(2,\mathbb{C})^{(b)} \oplus sl(2,\mathbb{C})^{(c)}$ and $so(4,\mathbb{C})$ takes a trivial form in the magic Bell basis [29] representation: $|\Psi^+\rangle = -i/\sqrt{2}(|0,1\rangle_{bc} + |1,0\rangle_{bc})$, $|\Psi^-\rangle = -1/\sqrt{2}(|0,1\rangle_{bc} - |1,0\rangle_{bc})$, $|\Phi^-\rangle = -i/\sqrt{2}(|0,0\rangle_{bc} - |1,1\rangle_{bc})$, $|\Phi^+\rangle = 1/\sqrt{2}(|0,0\rangle_{bc} + |1,1\rangle_{bc})$. Corresponding unitary transformation matrix is

$$\mathbf{U_B} = \frac{1}{\sqrt{2}} \begin{pmatrix} 0 & i & i & 0 \\ 0 & -1 & 1 & 0 \\ i & 0 & 0 & -i \\ 1 & 0 & 0 & 1 \end{pmatrix}. \tag{10}$$

In the Bell basis a generator $\chi \in sl(2,\mathbb{C})^{(b)} \oplus sl(2,\mathbb{C})^{(c)}$ takes the form $\chi_\mathbf{B} = \mathbf{U_B} \chi \mathbf{U_B}^\dagger$. Twelve $sl(2,\mathbb{C})^{(b)} \oplus sl(2,\mathbb{C})^{(c)}$ generators ($\mathbf{1}^{(b)} \otimes \sigma_{x,y,z}^{(c)}$, $i\mathbf{1}^{(b)} \otimes \sigma_{x,y,z}^{(c)}$, $\sigma_{x,y,z}^{(b)} \otimes \mathbf{1}^{(c)}$, $i\sigma_{x,y,z}^{(b)} \otimes \mathbf{1}^{(c)}$) map onto a set of twelve complex-valued *antisymmetric* matrices spanning $so(4,\mathbf{C})$: $\mathbf{1}^{(b)} \otimes \sigma_z^{(c)} \to -\mathbf{1} \otimes \sigma_y$, $\sigma_x^{(b)} \otimes \mathbf{1}^{(c)} \to -\sigma_x \otimes \sigma_y$, $\sigma_y^{(b)} \otimes \mathbf{1}^{(c)} \to -\sigma_y \otimes \mathbf{1}_z$, $\sigma_z^{(b)} \otimes \mathbf{1}^{(c)} \to \sigma_z \otimes \sigma_y$, $\mathbf{1}^{(b)} \otimes \sigma_{x,y}^{(c)} \to -\sigma_y \otimes \sigma_{x,z}$. Note that vectors $\mathbf{Z}$ and $\mathbf{W}$ introduced in [18] are equal to $\mathbf{Z} = \mathbf{U_B} \mathbf{c}_0^{(1)}$, $\mathbf{W} = \mathbf{U_B} \mathbf{c}_1^{(1)}$. In the Bell Basis antisymmetric matrix $\mathbf{P}^{(1)}$ from equation (6) takes the form

$$\mathbf{P}_\mathbf{B}^{(1)} = \mathbf{U_B} \mathbf{P}^{(1)} \mathbf{U_B}^T. \tag{11}$$

Matrix elements $P_{B,i,j}^{(1)} = \sum_{m,k=1}^{4} U_{B,i,m} P_{m,k}^{(1)} U_{B,j,k}$ can be also expressed as $P_{B,i,j}^{(1)} = Z_i W_j - Z_j W_i$. Note that $\mathbf{P}_\mathbf{B}^{(1)}$ is also an antisymmetric matrix.

For an arbitrary local operator $\mathbf{V} \in SL(2,\mathbf{C})^{(b)} \otimes SL(2,\mathbf{C})^{(c)}$ inducing an action $\mathbf{c}_{0,1}^{(1)} \to \mathbf{c}_{0,1}^{(1)\prime} = \mathbf{V}\mathbf{c}_{0,1}^{(1)}$ matrix $\mathbf{P}^{(1)}$ transforms as $\mathbf{P}^{(1)} \to \mathbf{P}^{(1)\prime} = \mathbf{V} \mathbf{P}^{(1)} \mathbf{V}^T$. Since in the Bell basis operator $\mathbf{V}$ takes the form of canonical $SO(4,\mathbf{C})$ operator $\mathbf{V_B} = \mathbf{U_B} \mathbf{V} \mathbf{U_B}^\dagger$ obeying a standard relation $\mathbf{V_B}^T = \mathbf{V_B}^{-1}$, transformation of Plücker matrix $\mathbf{P}_\mathbf{B}^{(1)}$ takes an explicit form of an adjoint $SO(4,\mathbf{C})$ action

$$\mathbf{P}_B^{(1)} \to \mathbf{P}_B^{(1)\prime} = \mathbf{V_B} \, \mathbf{P}_B^{(1)} \, \mathbf{V_B}^{-1}. \tag{12}$$

Induced transformations of six-dimensional Bell-basis Plücker vector $\mathbf{p}_B$, defined in equation (7), generates 6-dimensional irreducible adjoint representation of $SO(4,\mathbf{C})$.

Since $SO(4,\mathbf{C})$ preserves dot product, for any two matrices $\mathbf{P}_B$ and $\mathbf{R}_B$, obeying transformation relation (12), we have $Tr(\mathbf{P}_B \mathbf{R}_B) = Tr(\mathbf{P}_B' \mathbf{R}_B')$. Since $Tr(\mathbf{P}_B \mathbf{R}_B) = -2\mathbf{p}_B \cdot \mathbf{r}_B = -2\sum_i p_{B,i} r_{B,i}$, the (standard Euclidean) dot product becomes an invariant $\mathbf{p}_B \cdot \mathbf{r}_B = \mathbf{p}_B' \cdot \mathbf{r}_B'$, in particular, $\mathbf{p}_B \cdot \mathbf{p}_B = \mathbf{p}_B' \cdot \mathbf{p}_B'$. Taking into account that both vectors $\mathbf{p}^{(1)}$ and $\mathbf{p}_B^{(1)}$ by design do not change under $SL(2,\mathbf{C})^{(a)}$ set of operations acting on qubit $a$, the dot product $\mathbf{p}_B^{(1)} \cdot \mathbf{p}_B^{(1)}$ represents a three-qubit $SL(2,\mathbf{C})^{(a)} \otimes SL(2,\mathbf{C})^{(b)} \otimes SL(2,\mathbf{C})^{(c)}$ polynomial invariant. Peter Lévay found that three-tangle [15] is equal to four times the absolute value of this invariant

$$\tau_{abc} = 4\left|\mathbf{p}_B^{(1)} \cdot \mathbf{p}_B^{(1)}\right| = 4\left[\left(\mathrm{Re}\left[\mathbf{p}_B^{(1)} \cdot \mathbf{p}_B^{(1)}\right]\right)^2 + \left(\mathrm{Im}\left[\mathbf{p}_B^{(1)} \cdot \mathbf{p}_B^{(1)}\right]\right)^2\right]^{1/2}. \tag{13}$$

Plücker Bell-basis vectors in equation (13) are elements of $\mathbf{C}^6$ space (satisfying additional constraint (8)). As described in the next section equation (13) can be rewritten in the form of equation involving only two real-valued vectors from $\mathbf{R}^3$ space (for each partition). These vectors obey transformation properties identical to Bloch vectors describing single qubit states. In the next section we will also clarify algebraic relation between three vectors $\mathbf{p}^{(1,2,3)}$ related to three possible partitions of three-qubit system and we show that partition invariance of three-tangle $\left|\mathbf{p}_B^{(1)} \cdot \mathbf{p}_B^{(1)}\right| = \left|\mathbf{p}_B^{(2)} \cdot \mathbf{p}_B^{(2)}\right| = \left|\mathbf{p}_B^{(3)} \cdot \mathbf{p}_B^{(3)}\right|$ reduces to a trivial relation between a set of vectors in $\mathbf{R}^3$ space.

The complex $SO(4,\mathsf{C})$ group does not preserve Hermitian norm $\langle \mathbf{p}_B | \mathbf{p}_B \rangle$. However, if $SL(2,\mathsf{C})^{(b)} \otimes SL(2,\mathsf{C})^{(c)}$ action is restricted to $SU(2)^{(b)} \otimes SU(2)^{(c)}$ then $SO(4,\mathsf{C})$ reduces to a real form $SO(4,\mathsf{R})$ and Hermitian norm $\langle \mathbf{p}_B | \mathbf{p}_B \rangle \equiv \langle \mathbf{p} | \mathbf{p} \rangle = \sum_i p_i^* p_i$ becomes a three-qubit local invariant quantity. It turns out that algebraically it reproduces the (squared) concurrence between qubits a and (bc) [18]

$$\tau_{a(bc)} = 4 Det \boldsymbol{\rho}_a \equiv 4 Det \boldsymbol{\rho}_{bc} = 4 \langle \mathbf{p}^{(1)} | \mathbf{p}^{(1)} \rangle = 4 \mathbf{p}^{*(1)} \cdot \mathbf{p}^{(1)}. \tag{14}$$

Equations (13) and (14) explicitly demonstrate a close relation between Plücker coordinates and entanglement properties of a three-qubit system. In this paper we use algebraic technique based on homomorphism between $SU(4)$ and $SO(6)$ to derive a new set of Plücker coordinates, which we call $\mathbf{q}$-vectors. These vectors are obtained by a special linear transformation of the original Plücker coordinates. Strictly speaking the set of three vectors $\mathbf{q}^{(1,2,3)}$, describing three possible qubit partitions consists of 18 complex numbers. However, due to a very simple inter-partition relation between $\mathbf{q}^{(1)}$, $\mathbf{q}^{(2)}$ and $\mathbf{q}^{(3)}$ all information about entanglement properties of the system is encoded in only three three-dimensional complex vectors. We denote these vectors as $\mathbf{A}$, $\mathbf{B}$ and $\mathbf{C}$. Under local single-qubit operations on qubits $a$, $b$ and $c$ vectors $\mathbf{A}$, $\mathbf{B}$ and $\mathbf{C}$ evolve exactly as if these vectors are single-qubit Bloch vectors. However, in contrast with Bloch vectors, pairwise coupling between qubits results in coupling between corresponding pair of vectors $\mathbf{A}$, $\mathbf{B}$ and $\mathbf{C}$. Interestingly, two-tangles $\tau_{(bc)}$, $\tau_{(ca)}$ and $\tau_{(ab)}$ can be identified with (squares) of magnitudes or imaginary parts $(\mathbf{A}_I)$ of these vectors, for example $\tau_{(bc)} = 8 \mathbf{A}_I \cdot \mathbf{A}_I$. We also found that spatial orientation of these three vectors determines which two-qubit coupling operators are required for entanglement control. We illustrate this technique by using Plücker $\mathbf{q}$-representation for finding efficient two-qubit coupling sequence for (i) transforming $W$ state to $GHZ$ state, (ii) transforming $W$ state to biseparable state and iii) transforming biseparable state to $GHZ$ state. These are non-trivial quantum state transformations that we shall show, are greatly aided by the introduction of geometric Bloch-type description using $\mathbf{q}$-variables.

## IV. Transformation of Plücker variables under a set of $SU(4)$ two-qubit entangling operations: introducing modified Plücker variables.

In this section we address the problem of transformation of Plücker variables $\mathbf{p}^{(1)}$, $\mathbf{p}^{(2)}$ and $\mathbf{p}^{(3)}$ under $SU(4)$ two-qubit operations acting on subsystems $(bc)$, $(ca)$ and $(ab)$ correspondingly. It happens that an answer to this problem also provides a key to establishing a relation between eighteen homogeneous polynomials $p_{1,2...6}^{(1,2,3)}$ corresponding to Plücker coordinates for all three partitions.

To address the problems of $SU(4)$ action on vectors $\mathbf{p}^{(1,2,3)}$ we will reformulate this problem at the level of $su(4)$ Lie-algebra. An arbitrary operator $\mathbf{U} = e^{-i\mathbf{H}} \in SU(4)$ can be generated by a continuous evolution $\mathbf{V}(t) = e^{-i\mathbf{H}t}$ with a time-like parameter $t$ changing from $0$ to $1$ such that $\mathbf{U} = \mathbf{V}(t=1)$. Matrix $\mathbf{H}$ is usually interpreted as a physical Hamiltonian; generator $i\mathbf{H} \in su(4)$. Let us denote matrix entries of the Hamiltonian of subsystem $(bc)$ as $\mathrm{H}_{n,k}^{(bc)}$, $n,k = 1,...,4$. Taking into account that $id/dt\, \mathbf{c}_{0,1}^{(1)} = \mathbf{H}^{(bc)} \mathbf{c}_{0,1}^{(1)}$, evolution of the of Plücker matrix $\mathbf{P}^{(1)}$ (equation (6)) is given by $id/dt\, P_{n,m}^{(1)} = H_{n,k}^{(bc)} P_{k,m}^{(1)} + H_{m,k}^{(bc)} P_{n,k}^{(1)}$. In matrix notations

$$id/dt\, \mathbf{P}^{(1)} = \mathbf{H}^{(bc)} \mathbf{P}^{(1)} + \mathbf{P}^{(1)} \mathbf{H}^{(bc)T} \tag{15}$$

Corresponding six-dimensional complex vector $\mathbf{p}^{(1)}$ defined in equation (7) satisfies the following evolution equation

$$id/dt\, \mathbf{p}^{(1)} = \tilde{\mathbf{H}}^{(bc)} \mathbf{p}^{(1)} \tag{16}$$

Hamiltonian $\tilde{\mathbf{H}}^{(bc)}$ is a $6\times 6$ matrix

$$\tilde{\mathbf{H}}^{(bc)} = \begin{pmatrix} (H_{11}^{(bc)}+H_{22}^{(bc)}) & H_{23}^{(bc)} & H_{24}^{(bc)} & -H_{13}^{(bc)} & -H_{14}^{(bc)} & 0 \\ H_{32}^{(bc)} & (H_{11}^{(bc)}+H_{33}^{(bc)}) & H_{34}^{(bc)} & H_{12}^{(bc)} & 0 & -H_{14} \\ H_{42}^{(bc)} & H_{43}^{(bc)} & (H_{11}^{(bc)}+H_{44}^{(bc)}) & 0 & H_{12}^{(bc)} & H_{13}^{(bc)} \\ -H_{31}^{(bc)} & H_{21}^{(bc)} & 0 & (H_{22}^{(bc)}+H_{33}^{(bc)}) & H_{34}^{(bc)} & -H_{24}^{(bc)} \\ -H_{41}^{(bc)} & 0 & H_{21}^{(bc)} & H_{43}^{(bc)} & (H_{22}^{(bc)}+H_{44}^{(bc)}) & H_{23}^{(bc)} \\ 0 & -H_{41}^{(bc)} & H_{31}^{(bc)} & -H_{42}^{(bc)} & H_{32}^{(bc)} & (H_{33}^{(bc)}+H_{44}^{(bc)}) \end{pmatrix}. \tag{17}$$

Lie algebra defined by equation (17) is a fifteen-dimensional (the number of independent real components of the Hermitian matrix $\mathbf{H}^{(bc)}$) subalgebra of thirty-five dimensional Lie algebra $su(6)$. This subalgebra is a six-dimensional representation of $su(4)$ generated by $\mathbf{H}^{(bc)}$ acting on vectors $\mathbf{c}_{0,1}^{(1)}$. If $\mathbf{H}$ is Hermitian then $\tilde{\mathbf{H}}$ is also Hermitian and $\langle \mathbf{p}^{(1)} | \mathbf{p}^{(1)} \rangle$=const. One can also verify that if $i\mathbf{H} \in sl(4,\mathbb{C})$ then $i\tilde{\mathbf{H}} \in sl(6,\mathbb{C})$ because $Tr(\tilde{\mathbf{H}}) = 3 Tr(\mathbf{H})$ from (17). As we discuss below in the case when $i\mathbf{H}$ matrices generate a set of $SU(4)$ unitary two-qubit transformations Lie algebra (17) is isomorphic to $so(6)$. However, to map (17) onto a canonical form whereby $so(6)$ is represented by real-values antisymmetric $6\times 6$ matrices one has to identify a proper similarity transform. To find this transformation we analyze additional invariant of dynamic equation (16) associated with Plücker identity, equation (8).

One can easily verify that $\tilde{\mathbf{H}}^T \mathbf{\Omega} + \mathbf{\Omega} \tilde{\mathbf{H}} = Tr(\mathbf{H})\, \mathbf{\Omega}$. For generators $i\mathbf{H}$ from $su(4)$ or $sl(4)$ algebras $Tr(\mathbf{H}) = 0$, consequently

$$\tilde{\mathbf{H}}^T \mathbf{\Omega} + \mathbf{\Omega}\, \tilde{\mathbf{H}} = 0. \tag{18}$$

By differentiating bilinear quadratic form $\mathbf{x}^T \mathbf{\Omega}\, \mathbf{y}$ using (16) and (18) one can immediately verify that $\mathbf{x}^T \mathbf{\Omega}\, \mathbf{y}$ is an invariant bilinear form characterizing the group of transformations generated by matrices defined by equation (17).

The diagonal form $\mathbf{\Omega}_D$ of the symmetric real-valued matrix $\mathbf{\Omega}$ contains three positive $(+1)$ diagonal elements and three negative elements $(-1)$. Therefore matrices $i\tilde{\mathbf{H}}$ span a subalgebra of $so(3,3,\mathbb{C})$ (a complexification of $so(3,3)$ pseudo-orthogonal Lie algebra). Algebra $so(3,3,\mathbb{C})$ is isomorphic to algebra $so(6,\mathbb{C})$ which comprises a real-valued $so(6,\mathbb{R})$ algebra as a maximal compact subgroup (for technical details see, for example, [30]). For a set of matrices defined by equation (17) we have $i\tilde{\mathbf{H}} \subset so(6,\mathbb{C}) \cap su(6) \equiv so(6,\mathbb{R})$. In other words, the six-dimensional representation of $su(4)$ algebra provided by equation (17) is isomorphic to $so(6)$ algebra. Notice that at the level of Lie groups we have homomorphism between $SU(4)$ and $SO(6)$ groups $SO(6) \equiv SU(4)/\mathbb{Z}_2$, analogous to a well-known homomorphism between $SU(2)$ and $SO(3)$ groups (see, for example, [25]).

In order to map generators (17) onto a canonical $so(6)$ form of real-values antisymmetric matrices we find unitary matrix $\mathbf{U}_{pq}$ such that $\mathbf{U}_{pq}^T \mathbf{\Omega} \mathbf{U}_{pq} = \mathbf{1}$. Then variables $p_i$ are expressed as linear combination of a new set of variables $q_i$ as follows:

$$\mathbf{p}^{(1,2,3)} = \mathbf{U}_{pq} \mathbf{q}^{(1,2,3)}; \quad \mathbf{q}^{(1,2,3)} = \mathbf{U}_{pq}^\dagger \mathbf{p}^{(1,2,3)}. \tag{19}$$

Matrix $\mathbf{U}_{pq}$ is given by

$$\mathbf{U}_{pq} = \frac{1}{\sqrt{2}} \begin{pmatrix} i & 1 & 0 & 0 & 0 & 0 \\ 0 & 0 & 0 & -1 & i & 0 \\ 0 & 0 & -i & 0 & 0 & 1 \\ 0 & 0 & i & 0 & 0 & 1 \\ 0 & 0 & 0 & 1 & i & 0 \\ -i & 1 & 0 & 0 & 0 & 0 \end{pmatrix}. \tag{20}$$

In the space of variables $q_i$ the invariant bilinear form $\mathbf{x}^T \mathbf{\Omega}\, \mathbf{y}$ reduces to a standard dot product

$$\mathbf{p}^T \mathbf{\Omega} \mathbf{p}' = \left(\mathbf{U}_{pq} \mathbf{q}\right)^T \mathbf{\Omega}\, \mathbf{U}_{pq} \mathbf{q}' = \mathbf{q} \cdot \mathbf{q}'. \tag{21}$$

Dynamic equation (16) takes the form

$$d/dt\, \mathbf{q}^{(1)} = -i\, \mathbf{U}_{pq}^\dagger \tilde{\mathbf{H}}^{(bc)} \mathbf{U}_{pq} \mathbf{q}^{(1)}. \tag{22}$$

As we explicitly show in the next section, matrices $-i\, \mathbf{U}_{pq}^\dagger \tilde{\mathbf{H}}^{(bc)} \mathbf{U}_{pq}$ are real antisymmetric matrices generating orthogonal $SO(6)$ rotations in six-dimensional space of complex vectors $\mathbf{q}^{(1)}$. This is a simple consequence of the fact that bilinear invariant form associated with dynamic equation (22) is a standard Euclidian dot product as described in equation (21).

The structure of $SO(6)$ group is directly related to the properties of original $SU(4)$ group [25-27, 30-32]. Consider the most physically important subgroup of $SU(4)$ - a group of local $SU(2) \otimes SU(2)$ transformations. The image of this subgroup in $SO(6)$ is a $SO(3) \times SO(3)$ block-diagonal subgroup, i.e.

a pair of $SO(3)$ matrices acting in three-dimensional subspaces $(q_1, q_2, q_3)$ and $(q_4, q_5, q_6)$. Next, there are nine two-qubit coupling generators $i\sigma_{x,y,z}^{(b)}\sigma_{x,y,z}^{(c)}$. These generators span Cartan complement of $su(2) \oplus su(2)$ subalgebra of $su(4)$ [23, 24]. These coupling terms take the form of nine $3\times 3$ off-diagonal submatrices of $so(6)$, representing Cartan complement of $so(3) \oplus so(3)$ subalgebra of $so(6)$. To demonstrate these properties explicitly we represent Hamiltonian $\mathbf{H}^{(bc)}$ as a sum of single-qubit operators $\sigma_{x,y,z}^{(b)}$, $\sigma_{x,y,z}^{(c)}$ and nine coupling terms $\sigma_{xx,xy...zz}^{(bc)}$. Then we substitute that expression in equation (22) and simplify it. To carry this calculation in a meaningful fashion one needs to establish a map between $su(4)$ and generators and $so(6)$ generators.

## V. $SO(6)$ dynamics of $\mathbf{q}$-variables: a general case of qubit-qubit coupling

First we arrange $su(4)$ generators in the form of an antisymmetric $6\times 6$ matrix, each element of this matrix contains one of 15 $su(4)$ generators (mathematical details for constructing such a matrix are presented in the Appendix).

$$\mathbf{t}^{(bc)} = i\mathbf{\tau}^{(bc)} = i\begin{pmatrix} 0 & -\sigma_z^{(b)} & \sigma_y^{(b)} & \sigma_{xx}^{(bc)} & \sigma_{xy}^{(bc)} & \sigma_{xz}^{(bc)} \\ \sigma_z^{(b)} & 0 & -\sigma_x^{(b)} & \sigma_{yx}^{(bc)} & \sigma_{yy}^{(bc)} & \sigma_{yz}^{(bc)} \\ -\sigma_y^{(b)} & \sigma_x^{(b)} & 0 & \sigma_{zx}^{(bc)} & \sigma_{zy}^{(bc)} & \sigma_{zz}^{(bc)} \\ -\sigma_{xx}^{(bc)} & -\sigma_{yx}^{(bc)} & -\sigma_{zx}^{(bc)} & 0 & -\sigma_z^{(c)} & \sigma_y^{(c)} \\ -\sigma_{xy}^{(bc)} & -\sigma_{yy}^{(bc)} & -\sigma_{zy}^{(bc)} & \sigma_z^{(c)} & 0 & -\sigma_x^{(c)} \\ -\sigma_{xz}^{(bc)} & -\sigma_{yz}^{(bc)} & -\sigma_{zz}^{(bc)} & -\sigma_y^{(c)} & \sigma_x^{(c)} & 0 \end{pmatrix}. \tag{23}$$

For convenience in equation (23) we also included a Hermitian form $\tau_{n,m}$ of generators $\mathbf{t}_{n,m}$. It turns out that commutation relations for operators $\mathbf{t}_{n,m}^{(bc)}$ take a closed analytical form

$$\left[\mathbf{t}_{n,m}, \mathbf{t}_{k,p}\right] = 2(\delta_{m,p}\mathbf{t}_{n,k} + \delta_{n,k}\mathbf{t}_{m,p} - \delta_{m,k}\mathbf{t}_{n,p} - \delta_{n,p}\mathbf{t}_{m,k}). \tag{24}$$

Equation (24) defines structure coefficients for $su(4)$ algebra. There coefficients are identical to structure coefficients of $so(6)$ algebra: chose 15 generators $\mathbf{l}_{n,m}$ of $so(6)$ algebra to be $6\times 6$ real antisymmetric matrices with matrix entries $(\mathbf{l}_{n,m})_{i,j} = -\delta_{i,n}\delta_{j,m} + \delta_{i,m}\delta_{j,n}$. Commutation relations for these matrices (multiplied by a factor of 2) exactly reproduce equation (24).

$$\left[2\mathbf{l}_{n,m}, 2\mathbf{l}_{k,p}\right] = 2(\delta_{m,p}2\mathbf{l}_{n,k} + \delta_{n,k}2\mathbf{l}_{m,p} - \delta_{m,k}2\mathbf{l}_{n,p} - \delta_{n,p}2\mathbf{l}_{m,k}). \tag{25}$$

Thereby we have Lie algebra isomorphism $\mathbf{t}_{n,m} \leftrightarrow 2\mathbf{l}_{n,m}$. Notice that $e^{2\pi \mathbf{l}_{m,k}} = \mathbf{1}$, while $e^{\pi \mathbf{t}_{m,k}} = -\mathbf{1}$, i.e. $SO(6)$ is isomorphic to a quotient group $SU(4)/Z_2$ (here $Z_2$ is a group consisting of numbers $1$ and $-1$).

Matrices $\boldsymbol{\tau}_{2,1}^{(bc)}$, $\boldsymbol{\tau}_{3,1}^{(bc)}$, $\boldsymbol{\tau}_{3,2}^{(bc)}$... $\boldsymbol{\tau}_{6,5}^{(bc)}$ defined in equation (23) form a complete orthogonal basis in the space of Hermitian $4\times 4$ matrices: $Tr\left(\boldsymbol{\tau}_{n,m}\boldsymbol{\tau}_{p,q}\right)=4\delta_{n,p}\delta_{m,q}$, $m<n$. Hamiltonian $\mathbf{H}^{(bc)}$ can be represented in the form of a sum over generators $\boldsymbol{\tau}_{n,m}^{(bc)}$ as follows

$$\mathbf{H}^{(bc)} = \sum_{n=2}^{6}\sum_{m=1}^{n-1} f_{n,m}^{(bc)}\boldsymbol{\tau}_{n,m}^{(bc)}. \tag{26}$$

Here

$$f_{n,m}^{(bc)} = 1/4\, Tr[\mathbf{H}^{(bc)}\boldsymbol{\tau}_{n,m}^{(bc)}]. \tag{27}$$

Matrices $\boldsymbol{\tau}_{n,m}^{(bc)}$ are Hermitian and for Hermitian Hamiltonians $\mathbf{H}^{(bc)}$ coefficients $f_{n,m}^{(bc)}$ are all real-valued. While for simplicity we assume that the sum in equation (26) runs only over $m<n$, one can formally expand definition (27) to arbitrary values of indices $m=1,2...6$ and $n=1,2...6$. Since according to definition (23) $\boldsymbol{\tau}_{n,m}^{(bc)}=-\boldsymbol{\tau}_{m,n}^{(bc)}$, we have $f_{n,m}=-f_{m,n}$ and $6\times 6$ matrix $\mathbf{f}$, consisting of elements $f_{n,m}$, represents an antisymmetric real-valued generator from $so(6)$ algebra $\mathbf{f}\in so(6)$.

Substituting expression (26) in equation (22) and performing a set of simplifications we obtain a standard form of evolution equation describing $SO(6)$ rotations

$$\frac{d}{dt}q_n^{(1,2,3)} = 2\sum_{m=1}^{6} f_{n,m}^{(bc,ca,ab)} q_m^{(1,2,3)}, \; n=1,2...6. \tag{28}$$

For the sake of clarity, we have explicitly included in equation (28) all three partitions.

Notice that local Hamiltonian of qubit $b$ appears only in the upper $3\times 3$ block of matrix $\mathbf{f}$: $\mathbf{H}^{(b)} = 1/2\sum_{n,m=1}^{3} f_{n,m}^{(bc)}\boldsymbol{\tau}_{n,m}^{(bc)}$. Local $SU(2)^{(b)}$ transformations on qubit $b$ affect only coefficients $q_1^{(1)}$, $q_2^{(1)}$ and $q_3^{(1)}$. Qubit $b$ transformations are generated by a group of $SO(3)$-rotations represented by a set of orthogonal $3\times 3$ matrices embedded in the upper-diagonal $3\times 3$ block of $SO(6)$ matrix. Similarly, $SU(2)^{(c)}$ acting on qubit $c$ are generated by lower $3\times 3$ block of matrix $\mathbf{f}$, $\mathbf{H}^{(c)} = 1/2\sum_{n,m=4}^{6} f_{n,m}^{(bc)}\boldsymbol{\tau}_{n,m}^{(bc)}$ such that transformations acting on qubit $c$ act only on $q_4^{(1)}$, $q_5^{(1)}$ and $q_6^{(1)}$ components of vector $\mathbf{q}^{(1)}$, leaving $q_1^{(1)}$, $q_2^{(1)}$ and $q_3^{(1)}$ intact. Apparently generators from these two blocks commute with each other, as do local Hamiltonians $\mathbf{H}^{(b)}$ and $\mathbf{H}^{(c)}$. Coupling operations between qubits $c$ and $b$ are represented by $3\times 3$ off-diagonal block of matrix $\mathbf{f}^{(bc)}$: $\mathbf{H}^{(bc)} = \sum_{n=4}^{6}\sum_{m=1}^{3} f_{n,m}^{(bc)}\boldsymbol{\tau}_{n,m}^{(bc)}$. In other words, under the $SU(4)\to SO(6)$ map the $SU(2)^{(b)}\otimes SU(2)^{(c)}$ subgroup of $SU(4)^{(bc)}$ takes the block-diagonal form of $SO(3)^{(b)}\times SO(3)^{(c)}$ subgroup of $SO(6)^{(b,c)}$.

Suppose that for the problem of entanglement control a certain transformation $\mathbf{R} \in SO(6)$ should be implemented in the space of Plücker $\mathbf{q}^{(1)}$-coordinates (22). The corresponding physical Hamiltonian $\mathbf{H}^{(bc)}$ which will drive this transformation can be obtained via Lie algebra isomorphism $\mathbf{t}_{n,m} \leftrightarrow 2\mathbf{l}_{n,m}$. If transformation $\mathbf{R}$ is represented in the exponential form $\mathbf{R} = \exp\left(\sum_{n,m} x_{n,m} \mathbf{l}_{n,m}\right)$ a desired physical transformation acting in the Hilbert space of qubits $b$ and $c$ is given by $\mathbf{U}_R = \exp\left(1/2 \sum_{n,m} x_{n,m} \mathbf{t}_{n,m}\right)$ where a set of coefficients $x_{n,m}$ is exactly the same as for matrix $\mathbf{R}$. The physical Hamiltonian driving the desired quantum transformation is $\mathbf{H}^{(bc)} = 1/2 \sum_{n,m} x_{n,m} \boldsymbol{\tau}_{n,m}$. If several sequential control operations are being implemented such that $\mathbf{R} = \mathbf{R}_n \mathbf{R}_{n-1} ... \mathbf{R}_1$ then due to Lie group homomorphism we have $\mathbf{U}_R = \mathbf{U}_{R_n} \mathbf{U}_{R_{n-1}} ... \mathbf{U}_{R_1}$. Such a situation is ubiquitous in control theory. For example, Cartan $KAK'$ representation [23, 24] of an arbitrary two-qubit operation requires a decomposition of control operation onto a product of a sequence of two-qubit local operation and a coupling operation followed again by two-qubit local transformation. Note that inverse operation of finding matrix $\mathbf{R}_U \in SO(6)$ for a given matrix $\mathbf{U} \in SO(6)$ does not necessarily require representing $\mathbf{U}$ in the exponential form because it can be achieved via algebraic technique involving Pfaffians, for example [33].

### VI. Bloch $SO(3)$ evolution equations and inter-partition relations for Plücker q-variables.

Let us first introduce notations adequately reflecting the fact that vector space of $\mathbf{q}^{(1)}$-coordinates is physically partitioned into two three-dimensional subspaces due to the fact that local operations generated by Hamiltonians $\mathbf{h}^{(b)}$ and $\mathbf{h}^{(c)}$ are completely decoupled being represented by a $3 \times 3$ sub-matrices in equation (28).

Three possible qubit arrangements, $a(bc)$, $b(ca)$ and $c(ab)$ produce three sets of six-dimensional complex vectors $\mathbf{q}^{(1,2,3)} = \left(q_1^{(1,2,3)}, q_2^{(1,2,3)} ..., q_6^{(1,2,3)}\right)$. Let us partition six-dimensional vectors $\mathbf{q}^{(1,2,3)}$ into pairs of three-dimensional vectors $\boldsymbol{\alpha}^{(1,2,3)}$ and $\boldsymbol{\beta}^{(1,2,3)}$:

$$\boldsymbol{\alpha}^{(1,2,3)} = \left(q_1^{(1,2,3)}, q_2^{(1,2,3)}, q_3^{(1,2,3)}\right), \quad \boldsymbol{\beta}^{(1,2,3)} = \left(q_4^{(1,2,3)}, q_5^{(1,2,3)}, q_6^{(1,2,3)}\right). \tag{29}$$

Plücker $\mathbf{q}$-vectors are represented as a direct sum $\mathbf{q}^{(s)} = \boldsymbol{\alpha}^{(s)} \oplus \boldsymbol{\beta}^{(s)}, s \in (1,2,3)$. Notations (29) facilitate deriving relations connecting Plücker variables for all three partitions.

Consider local operation $SU(2)^{(a)}$ on qubit $a$. Its action on Plücker $\mathbf{q}$-variables is represented by $SO(3)^{(a)}$ real $3 \times 3$ sub-matrix and it affects only two vectors, $\boldsymbol{\alpha}^{(3)}$ and $\boldsymbol{\beta}^{(2)}$. Evolution of these vectors is induced by one local Hamiltonian $\mathbf{h}^{(a)}$. Dynamic equation (28), written for partitions $3$ and $2$ reduces to a Bloch-type evolution equation for vectors $\boldsymbol{\alpha}^{(3)}$ and $\boldsymbol{\beta}^{(2)}$

$$\frac{d}{dt}\mathbf{x} = 2\boldsymbol{\Theta}^{(a)} \times \mathbf{x}; \quad \mathbf{x} = \boldsymbol{\alpha}^{(3)}, \boldsymbol{\beta}^{(2)}. \tag{30}$$

where real three-dimensional vector $\mathbf{\Theta}^{(a)}$ is defined via coefficients of matrix $\mathbf{f}$ appearing in equation (28) for partitions $c(ab)$ and $b(ca)$: $\mathbf{\Theta}^{(a)} = (f_{32}^{(ab)}, -f_{31}^{(ab)}, f_{21}^{(ab)}) \equiv (f_{54}^{(ca)}, -f_{64}^{(ca)}, f_{65}^{(ca)})$. Local Hamiltonian driving qubit $a$ is simply $\mathbf{h}^{(a)} = \mathbf{\Theta}^{(a)} \cdot \mathbf{\sigma}$. Notice that Bloch vector of qubit $a$, defined as $m_n^{(a)} = Tr(\mathbf{\rho}_a \sigma_n)$, $n = 1, 2, 3$, satisfies dynamic equation identical to equation (30): $\dot{\mathbf{m}}^{(a)} = 2\mathbf{\Theta}^{(a)} \times \mathbf{m}^{(a)}$. Since vectors $\mathbf{\alpha}^{(3)}$ and $\mathbf{\beta}^{(2)}$, undergo the same $SO(3)$ evolution one may expect that $\mathbf{\alpha}^{(3)}$ and $\mathbf{\beta}^{(2)}$ are algebraically related. Indeed, comparing these vectors expressed as polynomials in coefficients $c_{i,j,k}$ given by equations (5), (19), (20) we find that $\mathbf{\beta}^{(2)} = -i\mathbf{\alpha}^{(3)}$. This relation establishes a trivial linear dependence between vectors $\mathbf{\alpha}^{(3)}$ and $\mathbf{\beta}^{(2)}$ consistent, obviously, with dynamic equation (30). Likewise, for other configurations we get cyclic inter-partition relations

$$\mathbf{\beta}^{(1)} = -i\mathbf{\alpha}^{(2)}; \; \mathbf{\beta}^{(2)} = -i\mathbf{\alpha}^{(3)}; \; \mathbf{\beta}^{(3)} = -i\mathbf{\alpha}^{(1)} \Rightarrow \mathbf{q}^{(1)} = (\mathbf{\alpha}^{(1)}, -i\mathbf{\alpha}^{(2)}), \; \mathbf{q}^{(2)} = (\mathbf{\alpha}^{(2)}, -i\mathbf{\alpha}^{(3)}), \; \mathbf{q}^{(3)} = (\mathbf{\alpha}^{(3)}, -i\mathbf{\alpha}^{(1)}). \quad (31)$$

Evidently, the full set of three six-dimensional $\mathbf{q}$-vectors reduces to a set of three linearly independent three-dimensional complex vectors. For simplicity, one can take these vectors to be $\mathbf{\alpha}^{(1)}$, $\mathbf{\alpha}^{(2)}$ and $\mathbf{\alpha}^{(3)}$. Equations (31) establish a link between Plücker $\mathbf{q}$-vectors for all three qubit configurations.

Additional relation satisfied by Plücker variables is identity (8), which due to equation (21) takes the form $\mathbf{q} \cdot \mathbf{q} = 0$. Therefore, for all partitions $\mathbf{\alpha}^{(n)} \cdot \mathbf{\alpha}^{(n)} = -\mathbf{\beta}^{(n)} \cdot \mathbf{\beta}^{(n)}$, $n = 1, 2, 3$. Combining this relation with equation (31) we get

$$\mathbf{\alpha}^{(1)} \cdot \mathbf{\alpha}^{(1)} = \mathbf{\alpha}^{(2)} \cdot \mathbf{\alpha}^{(2)} = \mathbf{\alpha}^{(3)} \cdot \mathbf{\alpha}^{(3)} = -\mathbf{\beta}^{(1)} \cdot \mathbf{\beta}^{(1)} = -\mathbf{\beta}^{(2)} \cdot \mathbf{\beta}^{(2)} = -\mathbf{\beta}^{(3)} \cdot \mathbf{\beta}^{(3)} \qquad (32)$$

At this point we would like to introduce notations reflecting physical meaning of equations involving vectors $\mathbf{\alpha}^{(1)}$, $\mathbf{\alpha}^{(2)}$ and $\mathbf{\alpha}^{(3)}$ (vectors $\mathbf{\beta}^{(1,2,3)}$ are related to vectors $\mathbf{\alpha}^{(1,2,3)}$ via equation (31)). Local rotations of qubit $b$ are affecting vector $\mathbf{\alpha}^{(1)}$ in $\mathbf{q}^{(1)} = (\mathbf{\alpha}^{(1)}, \mathbf{\beta}^{(1)}) \equiv (\mathbf{\alpha}^{(1)}, -i\mathbf{\alpha}^{(2)})$ in partition (1) - $a(bc)$. Therefore, we denote it as $\mathbf{\alpha}^{(1)} = \mathbf{B}$. Next, vector $\mathbf{\alpha}^{(3)}$ in $\mathbf{q}^{(3)} = (\mathbf{\alpha}^{(3)}, \mathbf{\beta}^{(3)}) \equiv (\mathbf{\alpha}^{(3)}, -i\mathbf{\alpha}^{(1)})$ in partition (3) - $c(ab)$ is rotated by local operations on qubit $a$, so we denote $\mathbf{\alpha}^{(3)} = \mathbf{A}$. Finally, vector $\mathbf{\alpha}^{(2)}$ in $\mathbf{q}^{(2)} = (\mathbf{\alpha}^{(2)}, \mathbf{\beta}^{(2)}) \equiv (\mathbf{\alpha}^{(2)}, -i\mathbf{\alpha}^{(3)})$ in partition (2) - $b(ca)$ so we denote is as $\mathbf{\alpha}^{(2)} = \mathbf{C}$. These notations also adequately reflect coupling between qubits: $\mathbf{h}^{(bc)}$ couples vectors $\mathbf{B}$ and $\mathbf{C}$, $\mathbf{h}^{(ca)}$ couples $\mathbf{C}$ and $\mathbf{A}$ etc. In new notations equation (32) takes the form $\mathbf{A} \cdot \mathbf{A} = \mathbf{B} \cdot \mathbf{B} = \mathbf{C} \cdot \mathbf{C}$.

### VII. Relation between Plücker q-variables and entanglement parameters.

Let us establish connection between Plücker $\mathbf{q}$-variables and quantities characterizing entanglement in the system of three qubits. First, consider evolution of complex vector $\mathbf{B} \equiv \mathbf{\alpha}^{(1)}$ under local operations on all three qubits. Note that from (31), vector $\mathbf{\alpha}^{(1)}$ appears in $\mathbf{q}^{(1)}$ in partition (1) - $a(bc)$ and in $\mathbf{q}^{(3)}$ in partition (3) - $c(ab)$. Therefore, vector $\mathbf{B}$ does not change under local operations on qubit $a$ because Plücker vectors $\mathbf{p}^{(1)}$ as well as $\mathbf{q}^{(1)}$ do not change (by design) under the action of operations on qubit $a$ in partition (1). Similarly, vector $\mathbf{B}$ also is not affected by local operations acting on qubit $c$ in partition (3), as discussed above. Vector $\mathbf{B}$ changes its orientation under local operations on qubit $b$. However, these transformations, being represented by $SO(3)$ matrices, do not affect the dot product $\mathbf{B} \cdot \mathbf{B}$. Therefore $\mathbf{B} \cdot \mathbf{B}$ represents a polynomial three-qubit invariant. Indeed, a straightforward comparison of three-tangle $\tau_{abc}$ [15] and $\mathbf{B} \cdot \mathbf{B}$ reveals that $\tau_{abc} = 8|\mathbf{B} \cdot \mathbf{B}|$. In view of equality (32) we immediately have partition-independent definition of three-tangle

$$\tau_{abc} = 8|\mathbf{B} \cdot \mathbf{B}| = 8|\mathbf{C} \cdot \mathbf{C}| = 8|\mathbf{A} \cdot \mathbf{A}| \tag{33}$$

Now we evaluate (squared) concurrences $\tau_{a(bc)}, \tau_{c(ab)}, \tau_{b(ca)}$, equation (14). Since vectors $\mathbf{p}$ and $\mathbf{q}$ are related by unitary transformation (19) we have $\langle \mathbf{p} | \mathbf{p} \rangle = \langle \mathbf{q} | \mathbf{q} \rangle$ and equation (14) takes a partition-symmetric form

$$\tau_{a(bc)} = 4\mathbf{q}^{(1)*} \cdot \mathbf{q}^{(1)} = 4(\mathbf{B}^* \cdot \mathbf{B} + \mathbf{C}^* \cdot \mathbf{C})$$
$$\tau_{b(ca)} = 4\mathbf{q}^{(2)*} \cdot \mathbf{q}^{(2)} = 4(\mathbf{C}^* \cdot \mathbf{C} + \mathbf{A}^* \cdot \mathbf{A}) \tag{34}$$
$$\tau_{c(ab)} = 4\mathbf{q}^{(3)*} \cdot \mathbf{q}^{(3)} = 4(\mathbf{A}^* \cdot \mathbf{A} + \mathbf{B}^* \cdot \mathbf{B})$$

As one expects, (squared) concurrence $\tau_{a(bc)}$ does not change under the full set of $SU(4)^{(bc)}$ operations, which are represented as $SO(6,\mathsf{R})^{(bc)}$ rotations acting on vectors $\mathbf{q}^{(1)} = (\boldsymbol{\alpha}^{(1)}, \boldsymbol{\beta}^{(1)}) \equiv (\mathbf{B}, -i\mathbf{C})$.

In order to establish a relation between vectors $\mathbf{q}^{(1,2,3)}$ and Wootters two-tangles $\tau_{(bc),(ac),(ba)}$ we need to make one more algebraic transformation of vectors $\mathbf{A}, \mathbf{B}, \mathbf{C}$. An intuitive driving idea is to make dot product entering equation (33) purely real. For example, for vector $\mathbf{A}$ we have $\mathbf{A}^2 = \mathbf{A}_R^2 - \mathbf{A}_I^2 + 2i\mathbf{A}_R \cdot \mathbf{A}_I$, where $\mathbf{A}_R = \mathrm{Re}\,\mathbf{A}$ and $\mathbf{A}_I = \mathrm{Im}\,\mathbf{A}$. So, one would wish to orthogonalize vectors $\mathbf{A}_R$ and $\mathbf{A}_I$ such that $\mathbf{A}_R \cdot \mathbf{A}_I = 0$. This can be achieved by multiplying the wave function by a global phase. Alternatively, instead of a global phase one can introduce additional phase factor directly in the definition of variables $\mathcal{P}_{(n,m),(k,l)}^{(1,2,3)}$ in equation (5). In any case, such an operation does not require any actual physical transformation applied to the system.

Multiplication of wave function coefficients $c_{i,j,k}$ in equation (1) by a phase factor $e^{i\phi}$ results in modification of vectors $\mathbf{q}^{(1,2,3)}$ by a factor $e^{2i\phi}$. As a result vector $\mathbf{A}$ becomes $\mathcal{A} = \mathbf{A}e^{2i\phi}$,

$$\mathcal{A}_R \cdot \mathcal{A}_I = \cos(4\phi)\mathbf{A}_R \cdot \mathbf{A}_I + \frac{1}{2}\sin(4\phi)(\mathbf{A}_R^2 - \mathbf{A}_I^2) \tag{35}$$

For the phase

$$\phi = -1/4 \arctan\left(2\mathbf{A}_R \mathbf{A}_I (\mathbf{A}_R^2 - \mathbf{A}_I^2)^{-1}\right) \tag{36}$$

the dot product $\mathcal{A}_R \cdot \mathcal{A}_I$ vanishes. Alternatively, one can require the dot product $\mathcal{A} \cdot \mathcal{A} = e^{4i\phi}\mathbf{A} \cdot \mathbf{A}$ to be purely real by choosing the phase $\phi$ as $\phi = -1/4 \arg(\mathbf{A}^2)$, which is identical to equation (36). Interestingly, for the gauge $\phi = -1/4 \arg(\mathbf{A}^2)$ dot products $\mathcal{B}_I \cdot \mathcal{B}_R$ and $\mathcal{C}_I \cdot \mathcal{C}_R$ vanish as well: relation $\mathcal{A}_I \cdot \mathcal{A}_R = \mathcal{B}_I \cdot \mathcal{B}_R = \mathcal{C}_I \cdot \mathcal{C}_R = 0$ is a trivial consequence of the Plücker identity (8) which is equivalent to $\mathcal{A} \cdot \mathcal{A} = \mathcal{B} \cdot \mathcal{B} = \mathcal{C} \cdot \mathcal{C}$.

Assuming, by default, that phase transformation was implemented in such a way that $|\mathcal{A}_R| \geq |\mathcal{A}_I|$, $|\mathcal{B}_R| \geq |\mathcal{B}_I|$ and $|\mathcal{C}_R| \geq |\mathcal{C}_I|$ we have

$$\tau_{abc} = 8(\mathcal{A}_R \cdot \mathcal{A}_R - \mathcal{A}_I \cdot \mathcal{A}_I) = 8(\mathcal{B}_R \cdot \mathcal{B}_R - \mathcal{B}_I \cdot \mathcal{B}_I) = 8(\mathcal{C}_R \cdot \mathcal{C}_R - \mathcal{C}_I \cdot \mathcal{C}_I) \tag{37}$$

Concurrences (34) are given by

$$\begin{aligned} \tau_{a(bc)} &= 4(\mathcal{B}_R \cdot \mathcal{B}_R + \mathcal{B}_I \cdot \mathcal{B}_I + \mathcal{C}_R \cdot \mathcal{C}_R + \mathcal{C}_I \cdot \mathcal{C}_I) \\ \tau_{b(ac)} &= 4(\mathcal{A}_R \cdot \mathcal{A}_R + \mathcal{A}_I \cdot \mathcal{A}_I + \mathcal{C}_R \cdot \mathcal{C}_R + \mathcal{C}_I \cdot \mathcal{C}_I) \\ \tau_{c(ab)} &= 4(\mathcal{A}_R \cdot \mathcal{A}_R + \mathcal{A}_I \cdot \mathcal{A}_I + \mathcal{B}_R \cdot \mathcal{B}_R + \mathcal{B}_I \cdot \mathcal{B}_I) \end{aligned} \tag{38 a,b,c}$$

Next, we use Coffman-Kundu-Wootters relation [15] which reads

$$\tau_{abc} = \tau_{a(bc)} - \tau_{(ac)} - \tau_{(ab)} \tag{39}$$

Rearranging terms in equation (37) we can represent three-tangle as $\tau_{abc} = 4(\mathcal{B}_R \cdot \mathcal{B}_R - \mathcal{B}_I \cdot \mathcal{B}_I + \mathcal{C}_R \cdot \mathcal{C}_R - \mathcal{C}_I \cdot \mathcal{C}_I)$. Combining this equation with first equation in (38) and equation (39) we have

$$8(\mathcal{B}_I \cdot \mathcal{B}_I + \mathcal{C}_I \cdot \mathcal{C}_I) = \tau_{(ac)} + \tau_{(ab)} \tag{40}$$

Repeating this operation for other partitions we also get

$$\begin{aligned} 8(\mathcal{A}_I \cdot \mathcal{A}_I + \mathcal{C}_I \cdot \mathcal{C}_I) &= \tau_{(bc)} + \tau_{(ab)} \\ 8(\mathcal{B}_I \cdot \mathcal{B}_I + \mathcal{A}_I \cdot \mathcal{A}_I) &= \tau_{(ac)} + \tau_{(bc)} \end{aligned} \tag{41}$$

Solving the system of equations (40), (41) for $\tau_{(ab)}$ $\tau_{(ac)}$ and $\tau_{(bc)}$ we have

$$\tau_{(bc)} = 8\mathcal{A}_I \cdot \mathcal{A}_I, \; \tau_{(ac)} = 8\mathcal{B}_I \cdot \mathcal{B}_I, \; \tau_{(ab)} = 8\mathcal{C}_I \cdot \mathcal{C}_I \tag{42}$$

For real parts of vectors $\mathcal{A}_R, \mathcal{B}_R, \mathcal{C}_R$ we also get

$$8\mathcal{A}_R \cdot \mathcal{A}_R = (\tau_{abc} + \tau_{(bc)}), \; 8\mathcal{B}_R \cdot \mathcal{B}_R = (\tau_{abc} + \tau_{(ac)}), \; 8\mathcal{C}_R \cdot \mathcal{C}_R = (\tau_{abc} + \tau_{(ab)}) \tag{43}$$

One can also derive a gauge-independent form of equations (42). For example, for vector $\mathbf{A}$ we have $\mathcal{A}_R \cdot \mathcal{A}_R + \mathcal{A}_I \cdot \mathcal{A}_I \equiv \mathbf{A} \cdot \mathbf{A}^*$ and $\mathcal{A}_R \cdot \mathcal{A}_R - \mathcal{A}_I \cdot \mathcal{A}_I = |\mathbf{A} \cdot \mathbf{A}| = \left[\left(\mathbf{A}_R^2 - \mathbf{A}_I^2\right)^2 + \left(2\mathbf{A}_R \cdot \mathbf{A}_I\right)^2\right]^{1/2}$ such that $\mathcal{A}_I \cdot \mathcal{A}_I = 1/2(\mathbf{A} \cdot \mathbf{A}^* - |\mathbf{A} \cdot \mathbf{A}|)$ and $\mathcal{A}_R \cdot \mathcal{A}_R = 1/2(\mathbf{A} \cdot \mathbf{A}^* + |\mathbf{A} \cdot \mathbf{A}|)$. Equations (42) takes gauge-invariant form:

$$\tau_{(bc)} = \frac{1}{4}\left(\mathbf{A} \cdot \mathbf{A}^* - |\mathbf{A} \cdot \mathbf{A}|\right), \ \tau_{(ac)} = \frac{1}{4}\left(\mathbf{B} \cdot \mathbf{B}^* - |\mathbf{B} \cdot \mathbf{B}|\right), \ \tau_{(ab)} = \frac{1}{4}\left(\mathbf{C} \cdot \mathbf{C}^* - |\mathbf{C} \cdot \mathbf{C}|\right), \tag{44}$$

One can verify that gauge-invariant equations (33), (34) and (44) result in Coffman-Kundu-Wootters relation (39).

Let us summarize our main result in this section.

1) We find that complete set of Plücker coordinates comprised of variables calculated for all three partitions contains three pairs of real-values 3D vectors $\mathcal{A}_R, \mathcal{A}_I, \mathcal{B}_R, \mathcal{B}_I, \mathcal{C}_R, \mathcal{C}_I$.
2) Magnitudes of these vectors determine all three concurrences, two-tangles and three-tangle.
3) Under local $SU(2)$ operations these vectors behave exactly like Bloch vector for corresponding qubits. Coupling between qubits $b$ and $c$ causes mixing of vectors $\mathcal{B}_R, \mathcal{B}_I$ and $\mathcal{C}_I, \mathcal{C}_R$, coupling between $a$ and $c$ cause interaction mixing of $\mathcal{A}_R, \mathcal{C}_I$ and $\mathcal{A}_I, \mathcal{C}_R$, coupling between $a$ and $b$ mixes $\mathcal{A}_R, \mathcal{B}_I$ and $\mathcal{A}_I, \mathcal{B}_R$.

### VIII. Examples of Entanglement Control: Manipulation of W and Biseparable states

**(A) Transforming $|W\rangle$ state to $|GHZ\rangle$**

According to Dür, Vidal and Cirac's classification of entanglement $|W\rangle$ and $|GHZ\rangle$ states belong to two separate classes [34]. Any transformation between states in different entanglement classes requires qubit coupling. We will apply technique developed in the previous section to perform a set of transformations on $|W\rangle$ state. The goal is to find an efficient route for transforming $|W\rangle$ state to $|GHZ\rangle$ state [35]. Standard definition of W state reads

$$|W\rangle = \frac{1}{\sqrt{3}}\left(|0,0,1\rangle + |0,1,0\rangle + |1,0,0\rangle\right). \tag{45}$$

Plücker $\mathbf{q}$-vectors for $|W\rangle$ state are identical for all partitions for an apparent reason: state (45) is symmetric relative to all qubit swap operation,

$$\mathbf{q}_W^{(1,2,3)} = (i, -1, 0, 1, i, 0)/3\sqrt{2}. \tag{46}$$

Three-dimensional vectors $\mathbf{A}_W = \mathbf{B}_W = \mathbf{C}_W = (i, -1, 0)/3\sqrt{2}$.

Entanglement parameters, equations (42)-(48), for $|W\rangle$ state are

$$\begin{aligned}
\tau_{abc}^W &= 0, \\
\tau_{(ab)}^W &= \tau_{(ac)}^W = \tau_{(bc)}^W = 4/9, \\
\tau_{a(bc)}^W &= \tau_{b(ac)}^W = \tau_{c(ab)}^W = 8/9.
\end{aligned} \tag{47}$$

When defining $|GHZ\rangle$ state we introduce an additional phase factor $\phi = \exp(-i\pi/4)$ in order to make Plücker vectors $\mathbf{A}, \mathbf{B}, \mathbf{C}$ purely real.

$$|GHZ\rangle = \frac{1}{\sqrt{2}} e^{-i\pi/4} (|0,0,0\rangle + |1,1,1\rangle). \tag{48a}$$

Plücker $\mathbf{q}$-vectors for $|GHZ\rangle$ state are

$$\mathbf{q}_{GHZ}^{(1,2,3)} = (0,0,1,0,0,-i)/2\sqrt{2}. \tag{48b}$$

Three-dimensional vectors are $\mathbf{A}_{GHZ} = \mathbf{B}_{GHZ} = \mathbf{C}_{GHZ} = (0,0,1)/2\sqrt{2}$. Straightforward calculation of three-tangle, and two-tangles for $|GHZ\rangle$ state gives

$$\begin{aligned} \tau_{abc}^{GHZ} &= 1, \\ \tau_{(ab)}^{GHZ} &= \tau_{(ac)}^{GHZ} = \tau_{(bc)}^{GHZ} = 0, \\ \tau_{a(bc)}^{GHZ} &= \tau_{b(ac)}^{GHZ} = \tau_{c(ab)}^{GHZ} = 1. \end{aligned} \tag{49}$$

To design a transformation which will transform $|W\rangle$ state into $|GHZ\rangle$ state let us first consider partition $a(bc)$ and compare vectors $\mathbf{q}_W^{(1)} = (B_w, -iC_W) = (i,-1,0,1,i,0)/3\sqrt{2}$ and $\mathbf{q}_{GHZ}^{(1)} = (B_{GHZ}, -iC_{GHZ}) = (0,0,1,0,0,-i)/2\sqrt{2}$. Our resources in changing these vectors are i) local $SO(3)$ rotations of vectors $\mathbf{B}_W$ and $\mathbf{C}_W$, plus ii) coupling of vectors $\mathbf{B}_W$ and $\mathbf{C}_W$ by orthogonal rotations generated by a set of 9 coupling generators $so(6)$: $\mathbf{l}_{1,4}, \mathbf{l}_{1,5}, \ldots \mathbf{l}_{3,6}$.

Our first transformation is designed to couple first and fifth components of vector $\mathbf{q}_W^{(1)}$ in order to eliminate imaginary entry in the first position. The desired rotation is generated by operator $\exp(\alpha \mathbf{l}_{1,5})$, $\alpha = \pi/4$ which performs an orthogonal rotation in the 1-5 hyper-plane by angle $\alpha = \pi/4$. The rotation results in transformation of vector $\mathbf{q}_W$ into vector $\mathbf{q}_{W_1} = (0,-1,0,1,\sqrt{2}i,0)/3\sqrt{2}$ (we have denoted new state as $|W_1\rangle$. Vector $\mathbf{B}_W$ becomes $\mathbf{B}_{W_1} = (0,-1,0)/3\sqrt{2}$ and vector $\mathbf{C}_W$ becomes $\mathbf{C}_{W_1} = (i,-\sqrt{2},0)/3\sqrt{2}$. This operation apparently will not change $\tau_{a(bc)}$, equation (38a), but it will increase the three-tangle $\tau_{abc} = 8(\mathcal{B}_R \cdot \mathcal{B}_R - \mathcal{B}_I \cdot \mathcal{B}_I)$, equation (37). Technically, this rotation "kills" imaginary part of vector $\mathbf{B}_W$ and adds it to the real component $\mathbf{C}_W$ such that vector $\mathbf{C}_{W,R}$ changes from $(0,-1,0)/3\sqrt{2}$ to $(0,-1,0)/3$. Let's clarify what happens with three- and two-tangles physically. According to equations (39), (42), (43) two-tangle $\tau_{(ac)}^W = 8\mathbf{B}_{W,I} \cdot \mathbf{B}_{W,I}$ and $(\tau_{abc}^W + \tau_{(ab)}^W) = (\tau_{a(bc)}^W - \tau_{(ac)}^W) = 8\mathbf{C}_{W,R} \cdot \mathbf{C}_{W,R}$. Two-tangle $\tau_{(ab)}^W = 8\mathbf{C}_{W,I} \cdot \mathbf{C}_{W,I}$ does not change because $\mathbf{C}_{W,I}$ does not change; $\tau_{a(bc)}$ also does not change because $\langle \mathbf{q}^{(1)} | \mathbf{q}^{(1)} \rangle$ does not change under $SO(6)^{(bc)}$ transformations. So, physically, while two-tangle $\tau_{(ac)}^W = 8\mathbf{B}_{W,I} \cdot \mathbf{B}_{W,I}$ disappears its value adds up to three-tangle which changes from $\tau_{abc}^W = 0$ to $\tau_{abc}^{W_1} = \tau_{abc}^W + \tau_{(ac)}^W = 4/9$.

Due to $SU(4) - SO(6)$ homomorphism and associated map between Lie algebra generators $\mathbf{t}_{n,m} \leftrightarrow 2\mathbf{l}_{n,m}$ one can immediately identify corresponding unitary $SU(4)^{(bc)}$ operator acting on the $|W\rangle$-state wave function which generates desired $SO(6)$ rotation in the $\mathbf{q}$-space. Namely, $SU(4)^{(bc)}$ operator corresponding to $SO(6)^{(bc)}$ operator $\exp(\alpha \mathbf{l}_{1,5}^{(bc)})$ is $\exp(\alpha/2 \mathbf{t}_{15}^{(bc)})$, $\mathbf{t}_{15}^{(bc)} = i\sigma_{xy}^{(bc)}$. New state $|W_1\rangle$ is equal to $\exp(i\pi/8\sigma_{xy}^{(bc)})|W\rangle$, explicitly,

$$|W_1\rangle = 1/\sqrt{6}\left(\sqrt{2-\sqrt{2}}|0,0,1\rangle + \sqrt{2+\sqrt{2}}|0,1,0\rangle + \sqrt{1+1/\sqrt{2}}|1,0,0\rangle - \sqrt{1-1/\sqrt{2}}|1,1,1\rangle\right). \tag{50}$$

Here $|W_1\rangle$ state is the state obtained after first transformation. Below we use $|W_n\rangle$ to denote a state after "n"-th transformation.

Next, we couple components 2 and 4 of vector $\mathbf{q}_{W_1}^{(1)} = (0,-1,0,1,\sqrt{2}i,0)/3\sqrt{2}$ by 2-4 rotation $\exp(\pi/4 \mathbf{l}_{2,4})$ in order to obtain Plücker $\mathbf{q}$-vector $\mathbf{q}_{W_2}^{(1)} = (0,-1,0,0,i,0)/3$. This operation kills imaginary part of vector $\mathbf{C}_{W_1}$ and adds it to real part of $\mathbf{B}_{W_1}$. Physically, $\tau_{(bc)}^{W_1}$ vanishes being transformed to three-tangle which will change from $\tau_{abc}^{W_1} = 4/9$ to $\tau_{abc}^{W_2} = \tau_{abc}^{W_1} + \tau_{(ab)}^{W_1} = 8/9$. Corresponding quantum operator is $\exp(\pi/8 \mathbf{t}_{24}^{(bc)}) = \exp(i\pi/8\sigma_{yx}^{(bc)})$. New state has the form

$$|W_2\rangle = \exp(i\pi/8\sigma_{yx}^{(bc)})|W_1\rangle = \frac{1}{\sqrt{6}}\left(|001\rangle + |010\rangle + \sqrt{2}|100\rangle - \sqrt{2}|111\rangle\right). \tag{51}$$

Entanglement parameters for this state are

$$\begin{aligned}
&\tau_{abc}^{W_2} = 8/9, \\
&\tau_{(ac)}^{W_2} = \tau_{(ab)}^{W_2} = 0, \tau_{(bc)}^{W_2} = 1/9, \\
&\tau_{a(bc)}^{W_2} = 8/9, \tau_{b(ac)}^{W_2} = \tau_{c(ab)}^{W_2} = 1.
\end{aligned} \tag{52}$$

By making two coupling transformations we achieved that vectors $\mathbf{q}_{W_2}^{(1)} = (0,-1,0,0,i,0)/3$ becomes similar to $\mathbf{q}_{GHZ}^{(1)} = (0,0,1,0,0,-i)/2\sqrt{2}$. Importantly, we have identified these two transformations simply by comparing vectors $\mathbf{q}_W^{(1)}$ and $\mathbf{q}_{GHZ}^{(1)}$.

There are no more coupling transformation on qubits $b$ and $c$ which can be implemented to transform $|W_2\rangle$ to $|GHZ\rangle$. We have to arrange transformations coupling qubits $a$ an $c$ or $a$ and $b$. Full set of Plücker $\mathbf{q}$-vectors for $|W_2\rangle$ state are

$$\begin{aligned}
\mathbf{q}_{W_2}^{(1)} &= \frac{1}{3}(0,-1,0,0,i,0), \\
\mathbf{q}_{W_2}^{(2)} &= \frac{1}{6\sqrt{2}}(0,-2\sqrt{2},0,1,3i,0), \\
\mathbf{q}_{W_2}^{(3)} &= \frac{1}{6\sqrt{2}}(i,-3,0,0,i2\sqrt{2},0).
\end{aligned} \tag{53}$$

Let us compare $\mathbf{q}_{W_2}^{(2)} = \left(0, -2\sqrt{2}/3, 0, 1/3, i, 0\right)/2\sqrt{2}$ and $\mathbf{q}_{GHZ}^{(2)} = (0,0,1,0,0,-i)/2\sqrt{2}$. By the same logic, one needs a rotation coupling components 2 and 4 of vector $\mathbf{q}_{W_2}^{(2)}$ in order to modify this vector into $\mathbf{q}_{W_3}^{(2)} = (0,-1,0,0,i,0)/2\sqrt{2}$. Desired rotation is provided by operator $\exp(\xi \mathbf{l}^{(ca)})$, $\xi = \arctan(1/2\sqrt{2})$. Corresponding physical coupling operator is $\exp(\xi/2 \mathbf{t}_{24}^{(ca)}) = \exp(\xi/2 \sigma_{yx}^{(ca)})$. The state $|W_2\rangle$ is transformed into

$$|W_3\rangle = \exp(\xi/2\sigma_{yx}^{(ca)})|W_2\rangle = \frac{1}{2}\left(|001\rangle + |010\rangle + |100\rangle - |111\rangle\right). \tag{54}$$

For the state $|W_3\rangle$ we have $\mathbf{q}_{W_3}^{(1,2,3)} = (0,-1,0,0,i,0)/2\sqrt{2}$. Entanglement parameters for the state $|W_3\rangle$ are identical to entanglement parameters of $GHZ$ state, equation (49). However, state $|W_3\rangle$ does not look like $|GHZ\rangle$ state. To make final adjustment let us compare vectors $\mathbf{A}_{W_3} = (0,-1,0)/2\sqrt{2}$ and $\mathbf{A}_{GHZ} = (0,0,1)/2\sqrt{2}$. Apparently one needs to perform a rotation $\exp(-\pi/2 \mathbf{l}_{23}^{(a)})$ which is represented by local operator $\exp(-\pi/4 \mathbf{t}_{56}^{(ca)})$ or $\exp(-\pi/4 \mathbf{t}_{23}^{(ab)}) = \exp(-i\pi/4 \sigma_x^a)$. This operation needs to be applied to vectors $\mathbf{B}_{W_3}$ and $\mathbf{C}_{W_3}$ as well, i.e. operator $\exp(-i\pi/4\sigma_x)$ is applied to all three qubits. The state becomes

$$|W_4\rangle = \exp\left[i\pi/4\left(\sigma_x^{(a)} + \sigma_x^{(b)} + \sigma_x^{(c)}\right)\right]|W_3\rangle = \frac{1}{\sqrt{2}}\left(i|000\rangle - |111\rangle\right), \tag{55}$$

For this state we have $\mathbf{q}_{W_4}^{(1,2,3)} = \mathbf{q}_{GHZ}^{(1,2,3)} = (0,1,0,0,-i,0)/2\sqrt{2}$. Trivial phase rotation $\exp(i\pi/4\sigma_z)$ and multiplication by a global phase factor $\exp(-i3\pi/4)$ result in canonical $|GHZ\rangle$ state.

$$|W_5\rangle = \exp(-i3\pi/4)\exp\left(i\pi/4\sigma_z^{(a)}\right)|W_4\rangle = \frac{1}{\sqrt{2}}(|000\rangle + |111\rangle). \tag{56}$$

In physical state space, we have realized the transformation $|W\rangle \rightarrow |W_5\rangle = |GHZ\rangle$ via the sequence of non-intuitive transformations, whose construction is greatly facilitated by working intuitively in $\mathbf{q}$-space:

$$|GHZ\rangle = \exp(-i3\pi/4)\exp\left(i\pi/4\sigma_z^{(a)}\right)\exp\left[i\pi/4\left(\sigma_x^{(a)} + \sigma_x^{(b)} + \sigma_x^{(c)}\right)\right]$$
$$\times \exp(\arctan(1/2\sqrt{2})/2\sigma_{yx}^{(ca)})\exp(i\pi/8\sigma_{yx}^{(bc)})\exp(i\pi/8\sigma_{xy}^{(bc)})|W\rangle.$$

**(B) Transforming a biseparable state to $|GHZ\rangle$**

Next example of state control is transformation of biseparable state to $|GHZ\rangle$ state. We define biseparable state as

$$|BS\rangle = \frac{1}{\sqrt{2}}(|000\rangle + |011\rangle) = |0\rangle \otimes \frac{1}{\sqrt{2}}(|00\rangle + |11\rangle). \qquad (57)$$

For the sake of clarity we calculate all three $\mathbf{q}$-vectors for this state:

$$\mathbf{q}_{BS}^{(1)} = (0,0,0,0,0,0),$$
$$\mathbf{q}_{BS}^{(2)} = (0,0,0,-1,-i,0)/2\sqrt{2}, \qquad (58)$$
$$\mathbf{q}_{BS}^{(3)} = (-i,1,0,0,0,0)/2\sqrt{2}.$$

Entanglement parameters are

$$\tau_{abc}^{BS} = 0,$$
$$\tau_{(ac)}^{BS} = \tau_{(ab)}^{BS} = 0, \tau_{(bc)}^{BS} = 1, \qquad (59)$$
$$\tau_{a(bc)}^{BS} = 0, \tau_{b(ac)}^{W_2} = \tau_{c(ab)}^{W_2} = 1.$$

We chose to perform transformations on vector $\mathbf{q}_{BS}^{(3)}$. Rotation $\exp(\alpha \mathbf{l}_{1,5}^{(ab)})$, $\alpha = \pi/4$, in the 1-6 hyperplane by angle $\pi/2$ transforms this vector to $\mathbf{q}_{BS_1}^{(3)} = (0,1,0,0,0,-i)/2\sqrt{2}$, apparently boosting the three tangle to its maximal possible value $\tau_{abc}^{BS_1} = 1$. Quantum transformation corresponding to this rotation is achieved by operator $\exp(i\pi/4\sigma_{xz}^{(ab)})$. The state becomes

$$|BS_1\rangle = \frac{1}{2}\left(|000\rangle + |011\rangle + \frac{i}{2}|100\rangle - \frac{i}{2}|111\rangle\right). \qquad (60)$$

New $\mathbf{q}$-vectors are

$$\mathbf{q}_{BS_1}^{(1)} = (0,0,1,0,0,-i)/2\sqrt{2},$$
$$\mathbf{q}_{BS_1}^{(2)} = (0,0,1,0,-i,0)/2\sqrt{2}, \qquad (61)$$
$$\mathbf{q}_{BS_1}^{(3)} = (0,1,0,0,0,-i)/2\sqrt{2}.$$

If we compare this set with the $\mathbf{q}$-set for GHZ state, $\mathbf{q}_{GHZ}^{(1,2,3)} = (0,0,1,0,0,-i)/2\sqrt{2}$, we see that the only difference is in $\mathbf{A}$-vectors. These vectors are defined as $\mathbf{A} = i(\mathbf{q}_4^{(2)}, \mathbf{q}_5^{(2)}, \mathbf{q}_6^{(2)}) \equiv (\mathbf{q}_1^{(3)}, \mathbf{q}_2^{(3)}, \mathbf{q}_3^{(3)})$ and for $|GHZ\rangle$ state we have $\mathbf{A}_{GHZ} = \mathbf{B}_{GHZ} = \mathbf{C}_{GHZ} = 1/2\sqrt{2}(0,0,1)$ while for $BS_1$ state we have $\mathbf{A}_{BS_1} = 1/2\sqrt{2}(0,1,0)$. Apparently we have to perform local $y-z$ rotation on qubit $a$. Corresponding operator is $\exp(\pi/4\mathbf{t}_{23}) = \exp(-i\pi/4\sigma_x^{(a)})$. The state becomes

$$|BS_2\rangle = \frac{1}{\sqrt{2}}(|000\rangle - i|111\rangle). \qquad (62)$$

Local $\exp(i3/4\pi\sigma_z)$ rotation applied to any qubit performs correction of the relative phase

$$|BS_3\rangle = \exp(i3/4\pi\sigma_z^{(a)})\frac{1}{\sqrt{2}}(|000\rangle - i|111\rangle) = \exp(i3/4\pi)\frac{1}{\sqrt{2}}(|000\rangle + |111\rangle). \tag{63}$$

The state $|BS_3\rangle$ apparently is a canonical $|GHZ\rangle$ state (up to a global $\exp(i3/4\pi)$ phase). Summarizing, we have transformed the biseparable state in (57) to the $|GHZ\rangle$ state $|BS\rangle \to |GHZ\rangle$ via the sequence of transformations in quantum state space

$$|GHZ\rangle = \exp(-i3/4\pi)\exp(i3/4\pi\sigma_z^{(a)})\exp(-i\pi/4\sigma_x^{(a)})\exp(i\pi/4\sigma_{xz}^{(ab)})|BS\rangle. \tag{64}$$

**(C) Transforming $|W\rangle$ state into a biseparable state**

It is interesting to note that $|W_2\rangle$ state, equation (51), obtained from $|W\rangle$ after two transformations $\exp(i\pi/8\sigma_{xy}^{(bc)})$ and $\exp(i\pi/8\sigma_{yx}^{(bc)})$, can be easily transformed into a biseparable state. To transform $|W_2\rangle$ to $|GHZ\rangle$ we have performed rotation of vector $\mathbf{q}_{W_2}^{(2)} = \frac{1}{6\sqrt{2}}(0,-2\sqrt{2},0,1,3i,0)$ in the plane $2-4$ by angle $\xi = \arctan(1/2\sqrt{2})$, instead we perform rotation by angle $\varsigma = -(\pi/2 - \xi)$ to get a state

$$|W_{BS}\rangle = \exp(\varsigma/2\sigma_{yx}^{(ca)})|W_2\rangle = \frac{1}{\sqrt{2}}(|001\rangle + |010\rangle) = |0\rangle \otimes \frac{1}{\sqrt{2}}(|01\rangle + |10\rangle). \tag{65}$$

Trivial qubit flip transformation $\sigma_x^{(b)}$ or $\sigma_x^{(c)}$ modified this state into biseparable state (57).

### IX. Summary and Discussion

In this section we briefly summarize conceptual elements of our work and discuss possible generalizations.

While the starting point of our work is the same as Peter Lévay's work [18], where he introduced Plücker coordinated in three-qubit problem, we have significantly extended the mathematical technique of Plücker description using $SO(6) - SU(4)$ homomorphism. As a result we have found concise partition-invariant description of entanglement in three-qubit system naturally tailored for geometric analysis of two-qubit coupling operations. This approach also revealed that from the point of view of quantum control theory entanglement parameters are represented not by a scalar quantities (three- and two-tangles) but by a set of Bloch-type three-dimensional vectors.

We introduce a new set of variables (quantum Plücker variables or $\mathbf{q}$-vectors) related to standard Plücker variables by a unitary rotation. Under this transformation quantum dynamics induced by two-qubit coupling takes the form of orthogonal $SO(6)$ group. We show that there exists a special global phase rotation which significantly simplifies relations between two-tangles and components of $\mathbf{q}$-vectors facilitating direct geometric interpretation of entanglement dynamics. We have identified that there exists an elegant relation between $\mathbf{q}$-vectors for all three partitions a(bc), c(ab) and b(ca) which allows to reduce redundant eighteen complex Plücker parameters to only three (complex) three-dimensional Bloch-type vectors. Under local rotations all three independent vectors $\mathbf{A}$, $\mathbf{B}$, $\mathbf{C}$ satisfy dynamic equations identical to standard equations for Bloch vectors (for each qubit), i.e. there exist six additional Bloch-type vectors: two real three-dimensional vectors for each qubit. Relative angles (and dot products) between these vectors and standard Bloch vectors are invariant under local rotations indicating that there may exist additional interesting and useful geometric properties of multipartite entanglement in three-qubit system.

As an illustration of our technique we apply it to a few quantum control problems: Geometric operations (rotations) of vectors $\mathbf{A}$, $\mathbf{B}$, $\mathbf{C}$ can be easily tailored to achieve a certain goal of transforming one state to another state. Using explicit analytical map between $SO(6)$ and $SU(4)$ groups we can establish protocol when desired quantum operations are mapped to 3D local rotations followed by qubit-qubit coupling in the form of coupling between vectors $\mathbf{A}$, $\mathbf{B}$ and $\mathbf{C}$. We have shown that the geometric $SO(6)$ description allows one to find an elegant and efficient solutions for these three problems, which would have been impossible to tackle in the original $SU(4)$ form.

We would like to note that there may not be any straightforward generalization of our method to four-qubit systems and non-pure quantum states since the chain of accidental isomorphism between su- and so- algebras ends at $su(4)$ algebra. However, we can generalize our approach to processes involving measurement operations via complexification of $su$- and $so$- algebras.

## Appendix: SU(4)-SO(6) Homomorphism

Consider an abstract set of elements $e_i$, $i = 1, 2 ..., n$, satisfying the following anticommutation relations $\{e_i, e_j\} = -2\delta_{i,j}$. Clifford algebra $Cl_n$ is a linear space generated by a set of $2^n - 1$ elements constructed as multiple products of first grade elements, i.e. $e_i$, $e_i e_j$, $e_i e_j e_k$ ... $e_1 e_2 e_3 ... e_n$. Due to anticommutation relations such an algebra is closed under multiplication i.e. a product of any two elements will always belong to one of $n$ grades of the algebra. It can be easily verified that a set of second-grade elements $e_i e_j$ form an $n(n-1)/2$-dimensional Lie algebra isomorphic to $so(n)$ Lie algebra with commutation relations identical to (23), (24).

$$[e_n e_m, e_k e_p] = 2(\delta_{n,k} e_m e_p + \delta_{m,p} e_n e_k - \delta_{n,p} e_m e_k - \delta_{m,k} e_n e_p). \tag{A.1}$$

The group generated by this algebra is called $Spin(n)$. It is a double cover of $SO(n)$ group. Explicitly, the adjoint action of $Spin(n)$ group on first-grade $n$-dimensional space algebraically reduces to an orthogonal rotation (for more details see [36]):

$$\exp\left(\sum_{i,j,i>j}^{n} x_{ij} e_i e_j\right) \sum_{i=1}^{n} \lambda_i e_i \exp\left(-\sum_{i,j,i>j}^{n} x_{ij} e_i e_j\right) = \sum_{i=1}^{n} \lambda'_i e_i. \tag{A.2}$$

Appearance of the adjoint action is a direct indication that $Spin(6)$ is a double cover of $SO(6)$.

Let's take a look at $Cl_6$ more closely. The number of elements in $Cl_6$ algebra is $2^n - 1 \equiv 63$ while the number of elements in the $spin(6)$ algebra is $n(n-1)/2 = 15$ (the same as the number of second-grade elements $e_i e_j$). Since the entire $Cl_6$ algebra has 63 elements it cannot be faithfully represented by 32-dimentional algebra of 4×4 complex matrices. To establishing a map between $su(4)$ and $so(6)$ Lie algebras one uses a compact representation of $spin(6)$ algebra.

Lie algebra $spin(6)$ is isomorphic to Lie algebra designed as a direct sum of second- and first-grade of $Cl_5$. Namely, commutation relations (A.1) hold if one replaces elements $e_6 e_1, e_6 e_2, e_6 e_3,\ldots$ by $e_1, e_2, e_3,\ldots$. $Cl_5$ has 31 elements and it can be faithfully represented by a set of 4×4 complex matrices. To establish a relation between $su(4)$ and $spin(6)$ we first construct a set of first grade elements $e_1, e_2,\ldots e_5$ of $Cl_5$ and then compute pairwise products $e_i e_j$ to generate the second-grade set. Then $\{e_i e_j\} \cup \{e_i\}$ will span a compact 15-dimentional algebra identical to the fundamental representation of $su(4)$ algebra.

First we need to identify elements of $su(4)$ which will serve as five first-grade elements of $Cl_5$ algebra. We have to choose a set of five elements out of 15 possible candidates: $i\sigma_{x,y,z}^{(b)} \otimes I^{(c)}$, $iI^{(b)} \otimes \sigma_{x,y,z}^{(c)}$, $i\sigma_{xx,xy\ldots}^{(bc)} \equiv i\sigma_{x,y,z}^{(b)} \otimes \sigma_{x,y,z}^{(c)}$. The following conditions have to be satisfied: 1) these five elements have to anti-commute $\{e_i, e_j\} = 0, for\, i \neq j$ and 2) pairwise products $e_i e_j$ should *not* belong to the set $e_i$ itself.

Lets' first include single-qubit Pauli matrices. We will chose $\mathbf{t}_{65} = e_5 = i\sigma_x^{(c)}$ and $\mathbf{t}_{64} = e_4 = -i\sigma_y^{(c)}$. Notice that operator $i\sigma_z^{(c)}$ will be included as a second-grade element $\mathbf{t}_{54} = e_5 e_4 = -i\sigma_x^{(c)} i\sigma_y^{(c)} = i\sigma_z^{(c)}$. We will complete the set of $e_{1,2,3,4,5}$ by three operators: $\mathbf{t}_{63} = e_3 = -i\sigma_{zz}^{(bc)}$, $\mathbf{t}_{62} = e_2 = -i\sigma_{yz}^{(bc)}$ and $\mathbf{t}_{61} = e_1 = -i\sigma_{xz}^{(bc)}$. Then all three conditions specified above are satisfied.

The rest of the task is trivial: by taking pairwise products of first-grade elements we recover the rest of the table in equation (23):

$$\begin{aligned}
\mathbf{t}_{21} &= e_2 e_1 = (-i\sigma_{yz}^{(bc)})(-i\sigma_{xz}^{(bc)}) = i\sigma_z^{(b)} \\
\mathbf{t}_{31} &= e_3 e_1 = (-i\sigma_{zz}^{(bc)})(-i\sigma_{xz}^{(bc)}) = -i\sigma_y^{(b)} \\
\mathbf{t}_{32} &= e_3 e_2 = (-i\sigma_{zz}^{(bc)})(-i\sigma_{yz}^{(bc)}) = i\sigma_x^{(b)} \\
\mathbf{t}_{41} &= e_4 e_1 = (-i\sigma_y^{(c)})(-i\sigma_{xz}^{(bc)}) = -i\sigma_{xx}^{(bc)} \\
\mathbf{t}_{42} &= e_4 e_2 = (-i\sigma_y^{(c)})(-i\sigma_{yz}^{(bc)}) = -i\sigma_{yx}^{(bc)} \\
\mathbf{t}_{43} &= e_4 e_3 = (-i\sigma_y^{(c)})(-i\sigma_{zz}^{(bc)}) = -i\sigma_{zx}^{(bc)} \\
\mathbf{t}_{51} &= e_5 e_1 = (i\sigma_x^{(c)})(-i\sigma_{xz}^{(bc)}) = -i\sigma_{xy}^{(bc)} \\
\mathbf{t}_{52} &= e_5 e_2 = (i\sigma_x^{(c)})(-i\sigma_{yz}^{(bc)}) = -i\sigma_{yx}^{(bc)} \\
\mathbf{t}_{53} &= e_5 e_3 = (i\sigma_x^{(c)})(-i\sigma_{zz}^{(bc)}) = -i\sigma_{zy}^{(bc)} \\
\mathbf{t}_{54} &= e_5 e_4 = (i\sigma_x^{(c)})(-i\sigma_y^{(c)}) = i\sigma_z^{(c)}
\end{aligned} \qquad (A.4)$$

By construction commutation relations for generators $\mathbf{t}_{i,j}$ are the same as generators $e_i e_j$ of $spin(6)$, given by equations (24), (A.1).

**Acknowledgements**
DU would like to thank support the Air Force Research Laboratory Summer Faculty Fellowship Program (AFRL-SFFP). PMA would like thank support from the Air Force Office of Scientific Research (AFOSR). Any opinions, findings, conclusions or recommendations expressed in this material are those of the author(s) and do not necessarily reflect the views of AFRL.